\documentclass[aps,superscriptaddress,nofootinbib,showpacs,eqsecnum,preprint,tightenlines]{revtex4}
\usepackage{hyperref}
\usepackage{epsfig,rotating}
\usepackage{amsmath,amssymb}
\usepackage{dsfont}
\usepackage{bbm}
\usepackage{slashed}
\numberwithin{equation}{section}

\newcommand{\be}{\begin{equation}}
\newcommand{\ee}{\end{equation}}
\newcommand{\bea}{\begin{eqnarray}}
\newcommand{\eea}{\end{eqnarray}}

\newcommand{\vx}{\vec{x}}

\newcommand{\vp}{\vec{p}}

\newcommand{\vq}{\vec{q}}

\newcommand{\vk}{\vec{k}}

\begin{document}

\title{Space-time evolution of heavy sterile neutrinos in cascade decays.}

\author{Daniel Boyanovsky}
\email{boyan@pitt.edu}
 \affiliation{Department of Physics and
Astronomy, University of Pittsburgh, Pittsburgh, PA 15260}

\date{\today}

\begin{abstract}
Heavy sterile-like neutrinos may be produced resonantly from the decay of pseudoscalar mesons and may decay into several different channels in a cascade $\Phi \rightarrow L^\alpha \nu_h;\nu_h\rightarrow \{X\}$. In general these are rare events with displaced vertices.   We provide a non-perturbative and manifestly unitary framework that describes the cascade decay and  yields the space-time evolution of  the probabilities for  sterile neutrinos, final states and the total number of events at a far detector. The results are general, valid for Dirac or Majorana neutrinos and only input the total decay rates and branching ratios for the production and decay channels.  We apply the  results to two examples of ``visible'' decay: i) $K^+\rightarrow e^+ \nu_h\rightarrow (e^+) e^+ e^- \nu_e$ via a standard model charged current vertex   and ii) the radiative decay $K^+\rightarrow \mu^+ \nu_h \rightarrow (\mu^+) \nu_a \gamma$. For this latter cascade process we find substantial corrections to previous assessments within  the parameter space argued to solve the anomalous excess of electron-like events at MiniBooNE. These large corrections  may help relieve the tension with recent experimental bounds on radiative decays  of heavy sterile neutrinos.

\end{abstract}

\maketitle

\section{Introduction}
Neutrino masses, mixing and oscillations are the clearest evidence yet of physics beyond the standard model \cite{book1,revbilenky,book2,book3}. They provide an explanation of the solar neutrino problem \cite{msw,book4,haxtonsolar} and have   important phenomenological \cite{book1,book2,book3,grimuslec,kayserlec,mohapatra,degouvea,bilenky}, astrophysical \cite{book4,book5,haxton} and cosmological \cite{dolgovcosmo} consequences. A wide range of experiments have confirmed  mixing and  oscillations among three light ``active'' neutrinos with $\delta m^2 = 10^{-4}-10^{-3}\,\mathrm{eV}^2$ for atmospheric and solar oscillations respectively.
Many extensions of the Standard Model that propose explanations via see-saw type mechanisms\cite{saw1,saw2,saw3,saw4} for neutrino masses predict the existence of heavy ``sterile'' neutrinos namely $SU(2)\times U(1)$ singlets that mix very weakly with ``active'' neutrinos. For a comprehensive discussion of the different scenarios see\cite{book1,book2,book3,mohapatra,book5}.  Heavy sterile neutrinos may play an important role in baryogenesis through leptogenesis\cite{yanagida,pila} or via neutrino oscillations\cite{akhruba} motivating several models for leptogenesis which may also yield dark matter candidates\cite{asa1,shapo,revs}. Furthermore, heavy sterile neutrinos may contribute to the energy transport during SNII explosions\cite{fullkuseSN}, their decay may be a source of early reionization\cite{earlyre}, they have been argued to play an  important role in the thermal history of the early Universe and to contribute to the cosmological neutrino background\cite{fullkuse}.  For a review of the role of sterile neutrinos in   cosmology and astrophysics  see ref.\cite{kuserev,revs,gorbunov}.

 Radiative decays of heavy sterile neutrinos via anomalous transition moments have been invoked as possible resolution of the LSND/MiniBooNE anomalies\cite{gninenko,gninenko2,dibfot} and another explanation of the LSND anomaly invokes the decay of a heavy sterile neutrino into light (active) neutrinos and scalars\cite{pascoli}.

  A comprehensive study of leptonic and semileptonic weak decays of heavy neutral leptons was carried out in ref.\cite{shrock} and extended in ref.\cite{rosner} and various experimental studies searching for heavy neutral leptons\cite{exp1,exp2,exp3,exp4,exp5,exp6,exp7,exp8,exp9,exp10,exp11,exp12,exp13,exp14,exp15} provide constraints on the values of the mixing matrix elements between heavy sterile and active neutrinos for a wide range of masses with   stringent bounds within the mass range $140\,\mathrm{MeV}\leq M_h \leq 500\,\mathrm{MeV}$\cite{exp12}. A summary of the bounds on the mixing matrix elements between sterile and active neutrinos is given in refs.\cite{gronau,kuseexpt2}.

   If the mass of the heavy sterile neutrino $m_h \lesssim M_{\pi,K},M_\tau$ they can be produced as resonances in the decay of pseudoscalar mesons  (or charged leptons)  opening a wide window for experimental searches. If heavy sterile neutrinos are Majorana, they can  mediate lepton number violating transitions with $|\Delta l| =2$ motivating further studies of their production and decay\cite{tao,dib,castro}.

The astrophysical, cosmological and phenomenological importance of heavy sterile neutrinos and their ubiquitous place in well motivated extensions beyond the Standard Model motivates   a series of recent proposals\cite{propo1,propo2,propo3,propo4,lou} that make a compelling case for rekindling the search for heavy sterile neutrinos in various current and next generation experiments.

 A thorough analysis of production and decay rates and cross sections\cite{shrock,rosner,tao,dib,propo1,propo2,propo3,propo4,lou} of heavy neutral leptons in various mass regimes provide the theoretical backbone for these proposed searches. Recent bounds on the mixing matrix elements between active (light) and sterile (heavy) neutrinos\cite{exp12,kuseexpt2} yield $|U_{eh}|^2;|U_{\mu h}|^2 \lesssim 10^{-7}-10^{-5}$ in the mass range $30 \,\mathrm{MeV} \lesssim m_h \lesssim 300 \, \mathrm{MeV}$ implying that the   production and decay rates of heavy neutrinos are exceedingly small, namely these are ``rare'' events. In particular the ``visible'' decay rates into charged leptons   being so small imply that  these processes result in displaced vertices and many of the proposed experiments envisage detectors placed far away from the production region. However, to the best of our knowledge, the study of the \emph{space-time} evolution of resonant sterile neutrinos from production to decay at a (far) detector  have not yet received the same level of attention.

\vspace{2mm}

\textbf{Motivation and goals:}
Motivated by the interest in  renewed searches for heavy neutral leptons in current and forthcoming experimental facilities, we explore a complementary aspect of the production and decay of heavy sterile neutrinos, namely the space-time evolution from the production to the decay region. Heavy neutral leptons of mass $m_h$ produced from the decay of pseudoscalar mesons ($ \Phi=\pi,K$) or charged leptons go on shell if $M_\Phi-m_L > m_h$ where $m_L$ is the mass of the charged lepton produced with the neutrino. This results in a resonant enhancement of the transition matrix element between the initial meson and final states from the decay of the heavy neutrino. For example the production of $\nu_h$ from pseudoscalar meson decay and the  decay of the heavy neutrino into a channel $\{X\}$ with invariant mass $m_X$  occurs, therefore in a sequential \emph{cascade} $\Phi \rightarrow L\,\nu_h ~;~ \nu_h \rightarrow \{X\}$, with a resonant transition matrix elements between initial and final states when $M_\Phi-m_L > m_h > m_X$.

The smallness of the mixing matrix elements between light (active) and heavy (sterile) neutrinos, imply that  the decay vertices $\nu_h \rightarrow \{X\}$ are far away from the production vertex and the number of decay events at a far detector will be influenced by the space time evolution of the heavy sterile neutrinos between the production and detection regions.

Our goal in this article is to study in detail this space time evolution   establishing a consistent formulation to assess the number of events measured at a far away detector. We seek to provide a general framework,  independent of the particular production and decay process so that an analysis of an experiment would  input the    branching ratios and decay rates that have been theoretically obtained in the literature into our results for the number of events at a far detector.

Our goal is thus different from previous efforts that focused on obtaining decay rates or branching ratios for particular processes, yet it is complementary in the sense that combining the results of our study for the space time evolution with the various production and decay rates available in the literature    yields a firmer understanding of the event rates and distributions at a far detector.

For this purpose we generalize and extend a recent study on the time evolution of cascade decay\cite{cascade} to the case of several production and decay channels. We combine this manifestly unitary framework with a wave-packet description to obtain the number of final state events detected at a far detector.

We apply this formulation to the study of two experimentally relevant cases of ``visible'' decay   analyzing in detail the consequences of the space-time evolution in these examples.

\section{The model.}

The total Hamiltonian is $H=H_0+H_I$ with $H_0$ the free field Hamiltonian and
\be H_I = H_M +H_{CC}+H_{NC}+H_{rad} \label{HI}\ee where
\be
H_M =F_{\Phi} \sum_{\alpha=e,\mu}\sum_j  \int d^3x \left[ U_{\alpha j} \,\overline{L}_{\alpha} (\vx,t)\, \gamma^{\mu} \mathbbm{P}_L\,\nu_j (\vx,t) (i   \partial_{\mu} \Phi (\vx,t)) \right] ~~;~~\mathbbm{P}_L =\frac{1}{2}(1-\gamma^5) \label{Hmeson}
\ee
 $\Phi$ is a complex (interpolating)  field that describes the charged pseudoscalar mesons $\Phi=\pi,K$. For a $\pi$ meson, we have that $F_{\pi} = \sqrt{2}\, G_F\, V_{ud} \, f_{\pi}$   and for the $K$ meson, we have that $F_{K} =  \sqrt{2}\, G_F V_{us}\, f_{K}$,  where $f_{\pi,K}$ are the decay constants, and $U$ is the  complex neutrino mixing matrix. The label $i$ in the sum runs over the (light) active-like $i\equiv a$  and the (heavy) sterile-like  $i\equiv h$ \emph{mass eigenstates}. $H_{CC},H_{NC}$ are the usual charged and neutral current vertices written in the neutrino mass basis.  Heavy sterile neutrinos may feature a non-vanishing transition magnetic moment\cite{book2} that allows for a radiative decay $\nu_h \rightarrow \nu_a\,\gamma$\cite{gninenko,gninenko2} this possibility is included in the interaction Hamiltonian via $H_{rad}$.

 $H_M$ describes the production of a (charged lepton) $L^\alpha$ and active (light) $\nu_a$ and  sterile-like (heavy) mass eigenstate $\nu_h$   from the decay of a  (pseudoscalar) meson $\Phi$  ($\Phi \rightarrow L^\alpha\,\nu_{a,h}$), $H_{CC}$   and   $H_{NC}$  describe  the  decay of the sterile-like heavy neutrino $\nu_h$ into a   multiparticle final state $\big\{X\big\}$ ($\nu_h \rightarrow \big\{X\big\}$) via Standard Model vertices in terms of mass eigenstates.   Non Standard Model couplings may be considered by including the corresponding terms in the interaction Hamiltonian in a straightforward generalization of the method described below.

\begin{figure}[h!]
\begin{center}
\includegraphics[height=4.5in,width=4.5in,keepaspectratio=true]{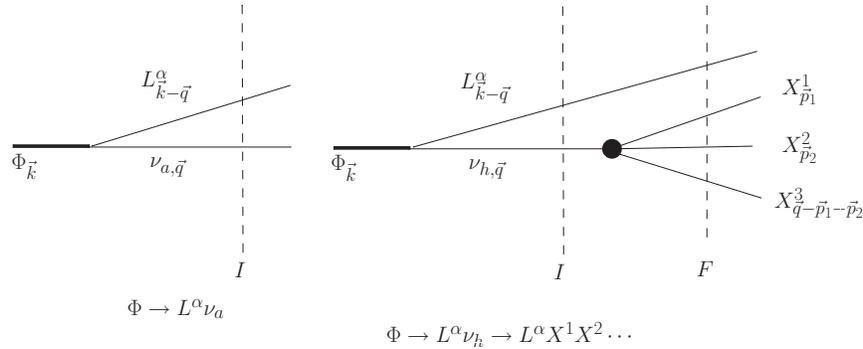}
\caption{Decay $\Phi \rightarrow L^\alpha\,\nu_a$ and cascade decay $\Phi \rightarrow L^\alpha\, \nu_h \rightarrow L^\alpha\,\big\{X\big\}$ where $\big\{X\big\}= X^1_{\vec{p}_1} X^2_{\vec{p}_2} X^3_{\vec{p}_3}\cdots$ is a multiparticle state with $\vec{p}_1+\vec{p}_2+\vec{p}_3+\cdots = \vec{q}$. The dashed lines depict the intermediate two particle state (I) and the final   multi  particle state (F).  }
\label{fig:cascade}
\end{center}
\end{figure}

 Let us consider an initial state   with one $\Phi$ particle of momentum $\vec{k}$ and the vacuum for the other fields,

 \be |\Psi(\vk,t=0)\rangle = |\Phi_{\vk}\rangle \,, \label{inistate} \ee upon time evolution this state evolves into $\big|\Psi(\vk,t)\rangle$ obeying
\be \frac{d}{dt} \big|\Psi(\vk,t)\rangle = -i (H_0+H_I) \big|\Psi(\vk,t)\rangle\,. \label{timeevol} \ee
When $M_\Phi > m_{L^\alpha}+m_{\nu_h}~;~m_{\nu_h} > m_{X}$ where $m_{X}$ is the invariant mass of the multiparticle state $\big\{X\big\}$, the interaction Hamiltonian (\ref{HI})  describes the cascade process depicted in fig.\ref{fig:cascade}.

Some examples of decay channels of the heavy (sterile-like) neutrino are
\be \nu_h\rightarrow \big\{X\big\} = \left\{
\begin{array}{c}
            e^+ e^- \nu_e \\
            \mu^+ \mu^- \nu_\mu \\
            e^- \mu^+ \nu_\mu \\
            \nu_a \gamma
          \end{array} \right.   ~~;~~ \nu_\alpha = \sum_a U_{\alpha\, a}\nu_a ~;~ \alpha = e,\mu \,. \label{decaychannels}
 \ee

We now pass to the interaction picture wherein
\be H_I(t) = H_I(t)= e^{iH_0t}\,H_I\,e^{-iH_0t} \label{HIint}\ee
and the state obeys
\be \label{inttimeevol}
i \frac{d}{dt} |\Psi(\vk,t)\rangle_I = H_I(t)|\Psi(\vk,t)\rangle_I
\ee
Consider that at $t=0$ the initial state is the single meson state of spatial momentum $\vk$ given by (\ref{inistate}),  at any later time, the  state $|\Psi(\vk,t)\rangle_I$ is   expanded in the basis of free particle Fock states $|n\rangle$ eigenstates of $H_0$, namely
\be |\Psi(\vk,t)\rangle_I = \sum_n A_n(t) |n\rangle \,. \label{ip}\ee

Up to second order in the interaction, the cascade decay depicted in fig. (\ref{fig:cascade}) is described by the following multiparticle state

\bea \label{intstate}
|\Psi(\vk,t)\rangle_I & = &  A_\Phi(\vec{k},t)\big|\Phi_{\vec{k}}\rangle + \sum_{\alpha;\vq;i=a,h}\,A^{\alpha\, i}_{I}(\vk,\vq;t)\,\big|\nu_{i,\vq};\,L^\alpha_{\vk-\vq} \rangle \nonumber \\ & + &    \sum_{\alpha;\vq;\{X\};\{\vec{p} \}_X} A^{\alpha\,X}_{F}(\vk,\vq,\{\vp\}_X;t)\,\big|L^\alpha_{\vk-\vq}\,; \{X\} \rangle  +\cdots
\eea

For simplicity of notation we do not distinguish between neutrino and antineutrino, furthermore,  the framework discussed below is general, independent of whether neutrinos are Dirac or Majorana.

In the last term in (\ref{intstate}), the sum over $\{X\}$ is over all the decay channels of $\nu_h$ and for each channel the sum over $\{\vp\}_X$ is over the momenta $\vp_1;\vp_2 \cdots $ of the multiparticle state $\{X\}$ constrained so that $\vec{p}_1+\vec{p}_2+\cdots = \vq$ (see fig.\ref{fig:cascade}). There is also an implicit sum over helicity states of the fermionic fields.
The coefficients $A_\Phi;A_I;A_F$ are the amplitudes of the initial, intermediate and final states, $\alpha = e,\mu$ are the charged leptons (we are considering either $\pi$ or $K$ decay but $\tau$ decay can be considered along the same lines as described below), each $\alpha$ represents a different decay channel for the pseudoscalar meson $\Phi$. The processes that lead to the state (\ref{intstate})  to second order in the interaction(s) are depicted in fig.(\ref{fig:cascade}), the dots stand for higher order processes, each vertex in the diagram (\ref{fig:cascade}) corresponds to one power of the couplings.

In perturbation theory there are also disconnected vacuum diagrams, these only renormalize the vacuum state and will not be considered here, a detailed discussion on these contributions is given in ref.\cite{boyrich}.

The amplitudes contain important information: the probability of finding a particular final state $X$ at time $t$ is given by $|A^{\alpha X}_F(t)|^2$ and the probability of finding a sterile neutrino in the intermediate state is $|A^{\alpha h}_I(t)|^2$. Furthermore, consider for example the final state from the decay $\nu_h \rightarrow e^+ e^- \nu_a$  the number of lepton \emph{pairs} in this state is given by (with the appropriate quantum numbers)
\be \langle \Psi(t)|b^\dagger_e b_e d^\dagger_e d_e|\Psi(t) \rangle = |A^{ e^+ e^- \nu_a}_F(t)|^2 \,, \label{dilenum}\ee and similarly, the number of sterile ($h$) and active ($a$) neutrinos in the intermediate state are respectively
\be \langle \Psi(t)|b^\dagger_h b_h |\Psi(t) \rangle = |A^{\alpha h}_I(t)|^2 ~~;~~ \langle \Psi(t)|b^\dagger_a b_a |\Psi(t) \rangle = |A^{\alpha a}_I(t)|^2\,.\label{hnum}\ee

The time evolution of the amplitudes $A_\Phi; A^i_I;A_F$ is obtained from the Schroedinger equation (\ref{inttimeevol}) by projecting onto the Fock states, namely with the interaction picture state written as (\ref{ip}) it follows that
 \be \dot{A}_m(t) = -i \sum_{n}\langle m|H_I(t)|n\rangle\,A_n(t)  = -i \sum_{n} \mathcal{M}_{mn}\, e^{i(E_m-E_n)t}\,A_n(t) \,. \label{eqamps1}\ee where we have used that the  matrix elements are of the form
\be
  \langle m | H_I(t) | n \rangle = e^{i(E_m-E_n)t}\,\mathcal{M}_{mn}~~;~~ \mathcal{M}_{mn} =\langle m|H_I(0)|n\rangle \,. \label{mtxele}
\ee

Since the interaction Hamiltonian is Hermitian unitary time evolution implies that
\be \langle \Psi(t)|\Psi(t)\rangle = |A_\Phi(\vec{k},t)|^2 + \sum_{\alpha;\vq;i=a,s}\,|A^{\alpha \,i}_{I}(\vk,\vq;t)|^2+\sum_{\alpha;\vq;\{X\};\{\vp \}_X} |A^{\alpha\,X}_{F}(\vk,\vq,\{\vp\};t)|^2 +\cdots = 1 \,. \label{unitti}\ee

It is clear from eqn. (\ref{eqamps1}) that the time evolution of the amplitudes is determined by a hierarchy of coupled equations which  will necessarily require a truncation to make any progress. Therefore confirming the unitarity relation (\ref{unitti}) in a final result   will be an important consistency check of the reliability of the results.

  We introduce the following notation,
\bea && E_\Phi \equiv E_{\Phi}(k)~~;~~E^i_I \equiv E_\alpha(|\vk-\vq|)+E_{i}(q)~~;~~ i = a, h\label{energies} \\&& E^X_F \equiv E_\alpha(|\vk-\vq|)+ E^X~~;~~E^X \equiv E_{X_1}(p_1)+E_{X_2}(p_2)+\cdots \label{defenergies}\\
&& \langle \nu_{i,\vq};\,L^\alpha_{\vk-\vq} |H_I(t)| \Phi_{\vk} \rangle \equiv M^{\alpha \,i}_\mathcal{P}(\vk,\vq)\,e^{-i(E_\Phi-E^i_I)t}      \label{Mpi}\\
&& \langle L^\alpha_{\vk-\vq}\,;\, \{X\} |H_I(t)|  \nu_{h,\vq};\,L^\alpha_{\vk-\vq} \rangle  \equiv M^{h\,X}_\mathcal{D}(\vk,\vq,\vp) \,e^{-i (E^i_I-E^X_F)t}     \label{Mdj}\eea where $E_\Phi(k);E_i(q);E_\alpha(|\vk-\vq|)$ are the single particle energies for the quanta of the respective fields and $E^X$ is the energy of the   multi-particle state with the set of momenta $\{\vp\}_X$. The matrix elements $M_\mathcal{P},M_\mathcal{D}$ refer to production ($\mathcal{P}$) and decay ($\mathcal{D}$) vertices with $ M^{\alpha\,i}_\mathcal{P} \propto U_{\alpha i}$. Only heavy sterile-like neutrinos feature a decay vertex.

To simplify notation    we  suppress  the momentum arguments of the amplitudes, energies and matrix elements, they are exhibited in the expansion (\ref{intstate}) and the definitions (\ref{energies},\ref{defenergies},\ref{Mpi},\ref{Mdj}) respectively.

Using eqn. (\ref{eqamps1}) we obtain the following equations for the amplitudes
\bea  \dot{A_{\Phi}}(t) && = -i \sum_{\alpha,\vq,i=a,h} {M^{\alpha\,i}_\mathcal{P}}^*\,e^{i(E_\Phi-E^i_I)t}\,A^{\alpha\,i}_I(t)
  ~;~ A_{\Phi}(0)=1  \label{dotafi1s}\\
 \dot{A^{\alpha\, a}_I}(t) && = -i\,e^{-i(E_\Phi-E^a_I)t}\,{M^{\alpha a}_\mathcal{P}} \, A_{\Phi}(t)~~;~~{A^{\alpha\,a}_I}(0)=0 \label{dotaI1a} \\
  \dot{A^{\alpha\,h}_I}(t) && = -i\,e^{-i(E_\Phi-E^h_I)t}\,{M^{\alpha h}_\mathcal{P}} \, A_{\Phi}(t) \nonumber \\ &&- i\sum_{\{X\};\{\vp\}_X} {M^{h\,X}_\mathcal{D}}^* \,e^{-i(E^X_F-E^h_I)t}\,A^{\alpha\,X}_F(t)  ~;~ {A^{\alpha\,h}_I}(0)=0 \label{dotaI1s} \\
  \dot{A^{\alpha X}_F}(t) && = -i   {M^{h\,X}_\mathcal{D}}  \,e^{i(E^X_F-E^h_I)t}\,A^{\alpha\,h}_I(t)~~;~~{A^{\alpha\,X} _F}(0)=0 \,,\label{dotaF1s}\eea the higher order terms in the expansion of the quantum state, represented by the dots in (\ref{intstate}) lead to higher order terms in the hierarchy of equations. It is shown in ref.\cite{cascade} that truncating the hierarchy at the order displayed above and solving the coupled set of equations provides a non-perturbative real time resummation of Dyson-type self-energy diagrams with self-energy corrections up to second order in the interactions and that the unitarity result (\ref{unitti}) is fulfilled up to the order considered. The reader is referred to (\cite{cascade}) for the technical details.

Equation (\ref{dotaI1s})  gives the evolution for the amplitude of the intermediate state with a resonant heavy sterile neutrino. It has a simple interpretation: the first term on the right hand side describes the \emph{production} of the sterile neutrinos ($\nu_h$) via the decay of the parent meson, the second term describes the \emph{decay} of ($\nu_h$) into the final state $X$.

 Since the active neutrinos $\nu_a$ do not decay we do not include their matrix elements that describe overlap with many particle states as these would be virtual leading only to a renormalization of the single particle energies.\footnote{It will become clear from the results obtained below that including the coupling of active (light) neutrinos to many particle states only leads to a renormalization of the single particle energy of the mass eigenstates.}.

The solution of the hierarchy of equations (\ref{dotafi1s}-\ref{dotaF1s}) is obtained by integrating from the bottom up. In the first step
\be A^{\alpha\,X}_F(t)= -i\, {M^{h\,X}_\mathcal{D}}  \,\int_0^t     \,e^{i(E^X_F-E^h_I)t'}\,A^{\alpha\,h}_I(t') \, dt'\, \label{afot1s}\ee inserting this solution into (\ref{dotaI1s}) we obtain
\be  \dot{A^{\alpha\,h}_I}(t)+ \sum_{{\{X\};\{\vp}\}_X}   \,\int_0^t   |M^{h X}_\mathcal{D}|^2 \,e^{i(E^h_I-E^X_F )(t-t')}\,A^{\alpha\,h}_I(t')\, dt'\,  =    -i\,e^{ i(E^h_I-E_\Phi )t}\,{M^{\alpha\,h}_\mathcal{P}}\, A_{\Phi}(t) \,. \label{dotaI11s}\ee

\vspace{2mm}

\textbf{The Wigner-Weisskopf approximation:} In solving the hierarchy of coupled equations from the bottom up, we encounter linear integro-differential equations for the coefficients, of the general form (see (\ref{dotaI11s})).

\be \dot{A}(t)+\int^t_0 \sum_{\vp}|M |^2 e^{i(E_I-E_F )(t-t')}\,A(t') dt' = I(t) \label{wweq}\ee where $I(t)$ is an inhomogeneity. These type of equations can be solved in terms of Laplace transforms (as befits an initial value problem). In ref.\cite{cascade} it is shown that the solution of  the hierarchy of equations via Laplace transform   yields a real time non-perturbative resummation of a Dyson-type self-energy diagrams. An alternative but equivalent method relies on that the matrix elements $M$ are typically of $\mathcal{O}(g)$ where   $g$ refers to a generic coupling in $H_I$. Therefore in perturbation theory the amplitudes evolve \emph{slowly} in time since $\dot{A} \propto g^2 A$ suggesting an expansion in \emph{derivatives}. This is implemented as follows\cite{cascade}, consider
\be \label{wzero}
W_0(t,t') = \sum_{\vp} |M|^2 \int^{t'}_{0} dt'' e^{-i(E_I-E_F )(t-t'')}
\ee which has the properties
\be
\frac{d}{dt'} W_0(t,t') =  \sum_{\vp} |M|^2 e^{-i(E_I-E_F )(t-t')}  \sim \mathcal{O}(g^2)~~;~~ W_0(t,0) = 0 \,. \label{propy}
\ee An integration by parts in (\ref{wweq}) yields

\be
\int^t_0 dt' \frac{d}{dt'} W_0(t,t') A (t') = W_0(t,t) A (t) - \int_0^t dt' \dot{A} (t') W_0(t,t') \label{deri}
\ee   From the amplitude equations it follows that  $\dot{A} \propto g^2\,A$ and $W_0 \propto g^2$, therefore the second term on the right hand side in (\ref{deri}) is $\propto g^4$ and can be neglected to leading order $\mathcal{O}(g^2)$ which is consistent with the order at which the hierarchy is truncated. This procedure can be repeated systematically, producing higher order derivatives, which are in turn higher order in $g^2$. The Wigner-Weisskopf approximation (to leading order) consists in keeping the first term in (\ref{deri}) and taking the long time limit,
 \be W_0(t,t) \rightarrow \sum_{\vp} |M|^2  \,\int^{t\rightarrow \infty}_0 e^{i(E_I-E_F +i \epsilon )(t-t'')} dt'' = i \sum_{\vp} \frac{|M|^2}{(E_I-E_F +i \epsilon )} \label{wwapx}\ee where $\epsilon \rightarrow 0^+$ is a convergence factor for the long time limit.

In ref.\cite{cascade} it is shown explicitly that this approximation is indeed equivalent to the exact solution via Laplace transform in the weak coupling and long time limit, where the Laplace transform is dominated by a narrow Breit-Wigner resonance  in the Dyson-resummed propagator.

Implementing the Wigner-Weisskopf approximation to leading order, eqn. (\ref{dotaI11s}) becomes
\be  \dot{A^h_I}(t)+ i\mathcal{E}_h \,A^h_I(t)   =    -i\,e^{ i(E^h_I-E_\Phi )t}\,{M^{\alpha h}_\mathcal{P}}^*\, A_{\Phi}(t) ~~;~~A^h_I(0)=0\,, \label{dotaI11sww}\ee where
\bea \mathcal{E}_h  & = &  \Delta E_h -i \frac{\Gamma_h}{2} \label{epsih} \\
\Delta E_h & = &   \sum_{\{X\}}\,  \sum_{\{\vp\}_X}  \,\mathcal{P}\,
\frac{|M^{h X}_\mathcal{D}|^2}{(E_h-E^X  )}    \label{deltaEh} \\
 \Gamma_h & = &  \sum_{\{X\}}\,\Big[2\pi \sum_{\{\vp\}_X}  {|M^{hX}_\mathcal{D}|^2}\,\delta(E_h-E^X  ) \Big] = \sum_{\{X\}}\,\Gamma(\nu_h \rightarrow \{X\}) \label{gammah}\,,  \eea where we used $E^h_I-E^X_F = E_h-E^X$ (see eqns. (\ref{energies},\ref{defenergies})) and $\Gamma(\nu_h \rightarrow \{X\})$ are the partial decay widths to the channel $\{X\}$. $\Gamma_h$ is the \emph{total} decay width of $\nu_h$, and   $\Delta E_h$ is absorbed  into a (mass) renormalization of $E_h$ ($E_h+\Delta E_h \rightarrow E_h$ where now $E_h$ is the renormalized energy).  The solution of (\ref{dotaI11sww}) is given by
\be A^{\alpha h}_I(t)  = -i{M^{\alpha h}_\mathcal{P}}^*\,e^{-i\mathcal{E}_h t}\,\int^t_0 e^{ i(E_h+E_\alpha-E_\Phi-i\frac{\Gamma_h}{2})t'} ~A_\Phi(t')\,dt'\,, \label{solAI1s}\ee in this expression $E_h$ is the renormalized energy of the sterile neutrino in the intermediate state. The solution of (\ref{dotaI1a}) is
\be  {A^{\alpha a}_I}(t) = -i\,{M^{\alpha a}_\mathcal{P}}^*\, \int^t_0e^{-i(E_\Phi-E^a_I)t'}\,  A_{\Phi}(t') \,dt' \,, \label{dotaI1asolu} \ee

Inserting the solutions (\ref{solAI1s},\ref{dotaI1asolu}) into eqn. (\ref{dotafi1s}) we obtain
\bea \dot{A}_\Phi(t) &+&  \sum_{\alpha;\vq} |M^{\alpha h}_\mathcal{P}|^2\,\int^t_0 e^{i(E_\Phi-E_h-E_\alpha+i\frac{\Gamma_h}{2})(t-t')} A_\Phi(t')\,dt' \nonumber \\ &+& \sum_{\alpha;\vq;a} |M^{\alpha a}_\mathcal{P}|^2\,\int^t_0 e^{i(E_\Phi-E_a-E_\alpha)(t-t')} A_\Phi(t')\,dt'   = 0 ~~;~~A_\Phi(0)=1 \,. \label{Afisol1s} \eea In the Wigner-Weisskopf approximation as in eqn. (\ref{dotaI11sww}) we obtain
\be \dot{A}_\Phi(t)+ i \mathcal{E}_\Phi\, A_\Phi(t) = 0 ~~;~~A_\Phi(0)=1 \,,\label{Afisol1sww} \ee with
\be \mathcal{E}_\Phi =  \sum_{\alpha;\vq} \frac{|M^{\alpha h}_\mathcal{P}|^2}{E_\Phi-E_h-E_\alpha +i \frac{\Gamma_h}{2}}+ \sum_{\alpha;\vq;a} \frac{|M^{\alpha a}_\mathcal{P}|^2}{E_\Phi-E_a-E_\alpha +i \epsilon } \equiv \Delta E_\Phi -i \frac{\Gamma_\Phi}{2}\,, \label{deltaEfi}\ee where

  \bea \Delta E_\Phi & = &   \sum_{\alpha;\vq}  \frac{|M^{\alpha h}_\mathcal{P}|^2\,\Big(E_\Phi-E_h-E_\alpha \Big)}{\Big[E_\Phi-E_h-E_\alpha \Big]^2+\Big[ \frac{\Gamma_h}{2}\Big]^2}+ \sum_{\alpha;a;\vq}  \mathcal{P}\, \frac{|M^{\alpha a}_\mathcal{P}|^2}{ E_\Phi-E_a-E_\alpha   }  \label{delfi}\\
  \Gamma_\Phi & = & 2\pi\,\sum_{\alpha;a;\vq}    {|M^{\alpha a}_\mathcal{P}|^2}\delta(E_\Phi-E_a-E_\alpha )+ \sum_{\alpha;\vq}  \frac{|M^{\alpha h}_\mathcal{P}|^2\,\Gamma_h}{\Big[E_\Phi-E_h-E_\alpha \Big]^2+\Big[ \frac{\Gamma_h}{2}\Big]^2}  \,.\label{deltaEfiww2} \eea $\Gamma_\Phi$ is the \emph{total} decay width of $\Phi$.

 The solution of (\ref{Afisol1sww}) is given by
\be A_\Phi(t) = e^{-i\Delta E_\Phi\,t}\,e^{-\frac{\Gamma_\Phi}{2}\,t}\,. \label{afifinww}\ee
 Going back to the Schroedinger picture the amplitude of the single $\Phi$ meson state becomes
 \be e^{-iE_\Phi t}A_\Phi(t) = e^{-i(E_\Phi+\Delta E_\Phi)t}\,e^{-\frac{\Gamma_\Phi}{2}\,t} \,, \label{ampsch}\ee the correction $\Delta E_\Phi$ is absorbed into a renormalization of the single particle energy of the $\Phi$ meson  $E_\Phi+\Delta E_\Phi \rightarrow E_\Phi$ with $E_\Phi$ now taken to be the renormalized single particle energy.

The first term in (\ref{deltaEfiww2}) is simply
 \be \sum_{\alpha;a}\,\Gamma(\Phi \rightarrow L^\alpha\,\nu_a)\,, \label{fiacti}\ee
 the second term  in (\ref{deltaEfiww2}) displays the resonant enhancement for the process $ \Phi \rightarrow L^\alpha\,\nu_h\rightarrow L^\alpha \{X\})$ and  becomes more familiar by writing    $\Gamma_h$  in the numerator using (\ref{gammah}), and writing in the narrow width limit
 \be \frac{1}{\Big[E_\Phi-E_h-E_\alpha \Big]^2+\Big[ \frac{\Gamma_h}{2}\Big]^2} \rightarrow \frac{2\pi}{\Gamma_h}\,\delta(E_\Phi-E_h-E_\alpha ) \label{narrow}\ee therefore, the second term in (\ref{deltaEfiww2}) becomes

\bea && \sum_{\alpha}\sum_{\{X\}}   \underbrace{  2\pi \sum_{\vq}|M^{\alpha h}_\mathcal{P}|^2 \delta(E_\Phi-E_h-E_\alpha )}_{\Gamma(\Phi \rightarrow L^\alpha \nu_h)}\, \underbrace{\Big[\frac{2\pi}{\Gamma_h} \sum_{\{\vp\}_X}  {|M^{hX}_\mathcal{D}|^2}\,\delta(E_h-E^X  ) \Big]}_{BR(\nu_h \rightarrow \{X\})}\nonumber \\ && = \sum_{\alpha}\,\sum_{\{X\}} \Gamma(\Phi \rightarrow L^\alpha \nu_h)\,BR(\nu_h \rightarrow \{X\})\,.\label{formula} \eea This is the familiar result for the decay rate via a resonant state in the narrow width approximation.

It remains to insert the solution (\ref{afifinww}) into (\ref{solAI1s}) and (\ref{dotaI1asolu}) leading to the following results
\be A^{\alpha h}_I(t) =  {M^{\alpha h}_\mathcal{P}} \,e^{-i\Delta E_h-\frac{\Gamma_h}{2}t}~
\frac{\Big[e^{-i(E_\Phi-E_h-E_\alpha-  \frac{i}{2}(\Gamma_\Phi-\Gamma_h))t}-1\Big]} {\Big[E_\Phi-E_h-E_\alpha -\frac{i}{2}(\Gamma_\Phi-\Gamma_h) \Big]}\,, \label{AIsfinit}\ee
\be A^{\alpha a}_I(t) =  {M^{\alpha a}_\mathcal{P}}\,e^{-i\Delta E_a t}~
\frac{\Big[e^{-i(E_\Phi-E_a-E_\alpha- \frac{i}{2} \Gamma_\Phi )t}-1\Big]} {\Big[E_\Phi-E_a-E_\alpha -\frac{i}{2} \Gamma_\Phi  \Big]}\,, \label{AIafinit}\ee this latter result is the same as that obtained in ref.\cite{lou} for the production of light active and sterile neutrinos from pseudoscalar decay.

Finally, we obtain the amplitude of the final state by inserting (\ref{AIsfinit}) into eqn. (\ref{afot1s}) obtaining (absorbing $\Delta E_h$ into the renormalized $E_h$)
\be A^{\alpha X}_F(t) = \frac{  {M^{\alpha h}_\mathcal{P}}  {M^{h X}_\mathcal{D}} }{E_\Phi-E_h-E_\alpha -\frac{i}{2}(\Gamma_\Phi-\Gamma_h)}
\Bigg[ \frac{ e^{-i(E_\Phi  - E_\alpha-E^X - i \frac{\Gamma_\Phi}{2})t} - 1}{E_\Phi  -E_\alpha- E^X - i \frac{\Gamma_\Phi}{2}} - \frac{e^{-i(E_h - E_X - i \frac{\Gamma_h}{2})t} -1 }{E_h  - E^X - i \frac{\Gamma_h}{2})} \Bigg]   \label{AFfinali}\ee

 From the final expressions for the amplitudes we now obtain the probability and total number of sterile neutrinos and specific final state particles. Let us first consider
 \be |A^{\alpha h}_I(\vk,\vq;t)|^2 = |{M^{\alpha h}_\mathcal{P}}|^2 \,e^{ - {\Gamma_h} t}~ \frac{|e^{ -i\mathcal{E}t}~e^{-\frac{\Delta \Gamma}{2}t}-1|^2}{\Big[\mathcal{E}^2+\Big( \frac{\Delta \Gamma}{2}\Big)^2 \Big]}~~;~~\mathcal{E} = E_\Phi-E_h-E_\alpha ~,~ \Delta \Gamma = \Gamma_\Phi-\Gamma_h \,. \label{probah}\ee In the narrow width limit  the above expression becomes $\propto \delta(\mathcal{E})$, to obtain the proportionality factor, we integrate it in the complex $\mathcal{E}$ plane wherein it features complex poles, we find
 \be |A^{\alpha h}_I(\vk,\vq;t)|^2 = 2\pi \,|{M^{\alpha h}_\mathcal{P}}|^2 \frac{\Big[e^{-\Gamma_h t}-e^{-\Gamma_\Phi t} \Big]}{\Gamma_\Phi-\Gamma_h}\,\delta(E_\Phi-E_h-E_\alpha)\,. \label{probahnu}\ee The \emph{total} number of sterile neutrinos produced from the decay of the pseudoscalar meson along with the charged lepton $L^\alpha$ is given by
 \be N^{(\alpha)}_{\nu_h}(t) = \sum_{\vq} |A^{\alpha h}(\vk,\vq;t)|^2 = \frac{\Big[e^{-\Gamma_h(q^*) t}-e^{-\Gamma_\Phi(k) t} \Big]}{\Gamma_\Phi(k)-\Gamma_h(q^*)}~\underbrace{2\pi \sum_{\vq}  |{M^{\alpha h}_\mathcal{P}}|^2\,\delta(E_\Phi-E_h-E_\alpha)}_{\Gamma(\Phi \rightarrow L^\alpha\nu_h)} \label{totnuh}\ee where $q^*$ is the value of the momentum of the sterile neutrino that satisfies energy conservation in the intermediate state, for a $\Phi$ meson decaying at rest $\vk =0$ and
 \be q^* = \frac{1}{2M_\Phi} \Big[M^4_\Phi+m^4_{L^\alpha}+m^4_{h}-2M^2_\Phi m^2_{L^\alpha}- 2M^2_\Phi m^2_{h}-2 m^2_{h}m^2_{L^\alpha}\Big]^{\frac{1}{2}} \,.\label{qstar}\ee Therefore we finally find
 \be N^{(\alpha)}_{\nu_h}(t) = \frac{\Big[e^{-\Gamma_h(q^*) t}-e^{-\Gamma_\Phi(k) t} \Big]}{1-\frac{\Gamma_h(q^*)}{\Gamma_\Phi(k)}}~BR(\Phi \rightarrow L^\alpha \nu_h)\,, \label{totnumnuh}\ee the superscript $(\alpha)$ refers to the fact that $\nu_h$ is kinematically entangled with the charged lepton $\alpha$ and $\Gamma_h(q^*)$  along with the branching ratio  depend on the mass of the particular charged lepton through the value of $q^*$.

 At early time $N^{(\alpha)}_{\nu_h}(t)$ grows as $\propto \big(\Gamma_\Phi(k)-\Gamma_h(q^*)\big)t$ clearly showing the production from meson decay minus the decay into the final products, and reaches a maximum at
 \be t^* = \frac{\ln\Big[\frac{\Gamma_\Phi(k)}{\Gamma_h(q^*)}\Big]}{\Gamma_\Phi(k)-\Gamma_h(q^*)}\,,\label{tmax}\ee after which it decays on the longer time scale\cite{cascade}.

  The \emph{total} number of \emph{active} neutrinos of species $a$ can be obtained from the above expression simply by setting $\Gamma_h=0$, namely
 \be N^{(\alpha)}_{\nu_a}(t) =  {\Big[1-e^{-\Gamma_\Phi(k) t} \Big]} ~BR(\Phi \rightarrow L^\alpha \nu_a) \label{totnumnua}\ee The calculation of $|A^X_F(t)|^2$ is more involved because of the many terms with interference among them, however the main steps have been discussed in ref.\cite{cascade}. Adapting the results from that reference  we find in the narrow width limit
 \bea |A^{\alpha X}_F(t)|^2   & = &   (2\pi)^2\,  \frac{|{M^{\alpha h}_\mathcal{P}}|^2\,  |{M^{h X}_\mathcal{D}}  |^2}{\Gamma_\Phi(k) \, \Gamma_h}\, \delta(E_\Phi-E_\alpha-E_h)\,\delta(E_h-E^X) \nonumber \\ & \times &
 \Bigg\{1-e^{-\Gamma_\Phi(k) t} - \frac{\Big[e^{-\Gamma_h t}-e^{-\Gamma_\Phi(k) t} \Big]}{1-\frac{\Gamma_h }{\Gamma_\Phi(k) }} \Bigg\} \label{AF2}\eea leading to
 \be \sum_{\vq,\{\vp\}_X}|A^{\alpha X}_F(t)|^2 = \Bigg\{1-e^{-\Gamma_\Phi(k) t} - \frac{\Big[e^{-\Gamma_h(q^*) t}-e^{-\Gamma_\Phi(k) t} \Big]}{1-\frac{\Gamma_h(q^*) }{\Gamma_\Phi(k) }} \Bigg\}\,BR(\Phi \rightarrow L^\alpha\,\nu_h)\,BR(\nu_h \rightarrow \{X\})\,. \label{sumita}\ee

Gathering the results (\ref{afifinww}, \ref{totnumnua},\ref{totnumnuh}, \ref{sumita}) we confirm the unitarity relation (\ref{unitti}). This is an important statement: the dependence of (\ref{sumita}) on \emph{both} the total decay widths of the parent meson and the sterile neutrino is \emph{a consequence of unitarity}. This observation will be important in the discussion of  experimentally relevant cases below.

 This expression simplifies in the case $\Gamma_\Phi \gg \Gamma_h$ and for $t \gg 1/\Gamma_\Phi$, namely well after the initial meson parent state has decayed

 \be \sum_{\vq,\{\vp\}_X}|A^{\alpha X}_F(t)|^2 \simeq  \Big\{1-e^{-\Gamma_h(q^*) t} \Big\}\,BR(\Phi \rightarrow L^\alpha\,\nu_h)\,BR(\nu_h \rightarrow \{X\})\,. \label{sumita2}\ee  We highlight that $\Gamma_h(q^*)$ is the \emph{total} decay width of the heavy sterile neutrino. For time scales $1/\Gamma_\Phi\ll t \ll 1/\Gamma_h(q^*)$ the above result simplifies further to
 \be \approx \Gamma^X_h(q^*)\,t\, BR(\Phi \rightarrow L^\alpha\,\nu_h)\label{shorti}\ee where $\Gamma^X_h(q^*)$ is the partial decay width of the heavy sterile neutrino into the specific channel $X$. This result has been invoked in the literature\cite{tao} but we emphasize that its validity is restricted to the case   $\Gamma_\Phi \gg \Gamma_h$ and time scales $1/\Gamma_\Phi \ll t \ll 1/\Gamma_h$.

 \vspace{2mm}

\textbf{Discussion}: the probabilities for finding active and heavy sterile neutrinos  and final states $X$ as a function of time are given respectively by
\bea  N^{(\alpha)}_{\nu_a}(t) & = &   {\Big[1-e^{-\Gamma_\Phi(k) t} \Big]} ~BR(\Phi \rightarrow L^\alpha \nu_a) \nonumber \\  N^{(\alpha)}_{\nu_h}(t) & = &  \frac{\Big[e^{-\Gamma_h(q^*) t}-e^{-\Gamma_\Phi(k) t} \Big]}{1-\frac{\Gamma_h(q^*)}{\Gamma_\Phi(k)}}~BR(\Phi \rightarrow L^\alpha \nu_h)\,,\label{Nnus} \eea

 \bea && N^{\alpha X}_F (t)   \equiv   \sum_{\vq,\{\vp\}_X}|A^{\alpha X}_F(t)|^2 \nonumber \\ & = &  \Bigg[\frac{\Gamma_\Phi(k)\,\Big(1-e^{-\Gamma_h(q^*) t}\Big) - \Gamma_h(q^*)\,\Big(1-e^{-\Gamma_\Phi(k) t} \Big)}{ \Gamma_\Phi(k)- \Gamma_h(q^*) } \Bigg]\,BR(\Phi \rightarrow L^\alpha\,\nu_h)\,BR(\nu_h \rightarrow \{X\})\,, \nonumber \\ &&  \label{NXs}\eea  which explicitly satisfy the unitarity relation (\ref{unitti}) up to the order at which the hierarchy has been truncated. These are some of the main results of this study. These results are broadly general they are valid either for Majorana or Dirac neutrinos and only depend on the total decay rates and branching ratios for the production and decay processes. In the case of Majorana neutrinos, these results also apply to lepton number violating $|\Delta l|=2$ transitions.

Although we cast the study in terms of production from the decay of parent mesons, the result is obviously more general and can be easily extended to the production of neutrinos from charged lepton decay with the obvious modification for additional particles in the intermediate state.

Oscillations between light active neutrinos (or light active and sterile) in the detection via charge current vertices can be straightforwardly analyzed as in ref.\cite{lou} by including another term in the Hamiltonian to describe the detection process via charged current interactions. We do not pursue this aspect here as our main interest is on the cascade decay process mediated by heavy sterile neutrinos.

The results above were obtained considering all particle states to be described by momentum eigenstates, namely plane waves. A space-time description of   production and decay is obtained by considering that the initial state is described by a spatially localized wave packet so that the initial (single meson) state is given by
\be |\Psi(\vk_0;t=0)\rangle = \sum_{\vk} C_\Phi(\vk,{\vk}_0)|\Phi_{\vk} \rangle, \label{wp}\ee where $C(\vk,\vk_0)$ is the spatial Fourier transform of the meson wave function and is sharply localized around the value $\vk_0$ corresponding to the average momentum of the meson wavepacket.

As a specific example we consider Gaussian wave packets normalized to unity within a volume $V$,
\be   C_\Phi(\vk,\vk_0) = \Bigg[\frac{8\,\pi^\frac{3}{2}}{\sigma^3\,V} \Bigg]^\frac{1}{2}~e^{-\frac{(\vk-\vk_0)^2}{2\sigma^2}}  \label{gaussianwf}\,,\ee where $\vk_0$ is the average momentum of the wavepacket and $\sigma$ is the localization scale in momentum space. The total number of mesons in the initial state is
\be N_\Phi = \sum_{\vk}\langle \Psi(\vk_0;t=0)\big|a^{\dagger}_{\vk,\Phi} a_{\vk,\Phi}\big|\Psi(\vk_0;t=0)\rangle = \sum_{\vk} | C_\Phi(\vk,\vk_0)|^2 =1 \,.\label{numfi}\ee

The spatial wave function is
\be F(\vx) = \Bigg[\frac{\sigma}{\sqrt{\pi}}\Bigg]^{3/2}\, e^{ i\vec{k}_0\cdot \vec{x}}~ e^{-\frac{1}{2}\,\sigma^2  \vec{x}^2} \,,\label{psigau}\ee it is localized at $\vec{x}=0$ with  localization length   $1/\sigma$.

 The linearity of time evolution implies that the same $C_\Phi(\vk,\vk_0)$ multiplies all the amplitudes in $|\Psi(t)\rangle$, this can be seen straightforwardly because now $A_\Phi(0) \rightarrow C_\Phi(\vk,\vk_0)\,A_\Phi(0)$ and the solutions of the evolution equations (\ref{dotafi1s})-(\ref{dotaF1s}) imply that the amplitudes for the intermediate and final states are also multiplied by this factor. Therefore in the Schroedinger picture the time evolved state is given by
 \be |\Psi(t)\rangle = \sum_{\vk} C_\Phi(\vk,\vk_0)\,|\Psi(\vk,t)\rangle_S \,\label{psikt}\ee where $|\Psi(\vk,t)\rangle_S$ is given by (\ref{intstate}) with $A_\Phi(t)\rightarrow e^{-iE_\Phi\,t}A_\Phi(t)~;~A_I(t) \rightarrow e^{-iE_I\,t}\,A_I(t)~;~A_F(t) \rightarrow e^{-iE_F\,t}$ and the amplitudes are the ones obtained above for the plane wave initial state. The sum (integral) over $\vk$ can now be done by expanding around $\vk = \vk_0$ as usual for localized wave packets, for example
 \be e^{-i E_\Phi(k)\,t} \, e^{-\Gamma_\Phi(k)\,t} \simeq  e^{-iE_\Phi(k_0)\,t}\,e^{-i \vec{v}_g(\vk_0)\cdot (\vk-\vk_0)\,t}\,e^{-\frac{1}{2}\Gamma_\Phi(k_0)\,t}\ee where we have used that $\Gamma_\Phi(k) = \frac{M_\Phi}{E_\Phi(k)}\Gamma_\Phi(0)$ and neglected terms $\mathcal{O}(\Gamma_\Phi/E_\Phi)\ll 1$ with similar expansions for the corresponding quantities for neutrinos.

  The spatial Fourier transform leads to a Gaussian wave function \be F(\vec{x},t) \propto e^{-\frac{\sigma^2}{2}(\vec{x}-\vec{v}_g\,t)^2}\,e^{-\frac{1}{2}\Gamma_\Phi(k_0)\,t}\,e^{ i\vec{k}_0\cdot \vec{x} -iE_\Phi(k_0)\,t} \,,\label{gausfina}\ee where we have neglected the  dispersion of the wave packet.

  Neutrinos produced from meson (or charged lepton) decay ``inherit'' the  wave packet profile through $C(\vk,\vec{k}_0)$ multiplying the amplitudes. Assuming that the spatial localization scale of the initial wave packet is smaller than the decay length of the parent particle (in the case of $\pi$ decay at rest this scale is $\sim 8\,\mathrm{mts}$) the spread in momentum of the wave packet around $\vk_0$ is larger than the decay width and the Lorentzian distributions of the decay products can be safely replaced by sharp delta functions as discussed above. Upon time evolution the spatial wave function of the  heavy sterile neutrinos features a maximum at $\vec{x}-\vec{v}^*_g\,t$ where $\vec{v}^*_g = \vec{q}^*(k_0)/E_\nu(q^*(k_0))$ is the group velocity   for   $q^*(k_0)$, which is the value of the sterile neutrino momentum that satisfies energy conservation for $\vec{k}=\vec{k}_0$ as discussed above. The   time dependent probabilities (\ref{Nnus},\ref{NXs}) are multiplied by $ e^{- {\sigma^2} (\vec{x}-\vec{v}^*_g\,t)^2}$ yielding a space-time description of the propagation. For a more extensive discussion on wave packets within a related framework see ref.\cite{louentan}.

  When the center of the wave packet has traveled a distance $L_d$, the probabilities (integrated in region of width $\sigma$ around the center of the wave packet) are given by (\ref{Nnus},\ref{NXs}) evaluated at $t = L_d/v^*_g$, furthermore since
 \be \Gamma_h(q^*) = \frac{m_h}{E_h(q^*)}\Gamma_h(0)\,,  \ee  the factors
  \be \Gamma_h(q^*)\, t = \Gamma_h(q^*)\,\frac{L_d}{v^*_g}= \frac{m_h}{q^*}\,   \frac{{L_d}}{\tau_h}  ~~;~~\Gamma_\Phi(k)\,t = \Gamma_\Phi(k)\,\frac{L_d}{v^*_g}=
  \frac{M_\Phi E_h(q^*)}{q^*E_\Phi(k)}\, \,\frac{L_d}{\tau_\Phi}\label{expo}\ee where $\tau_h,\tau_\Phi$ are the decay lifetimes at rest.

 Consider a detector of length $\Delta L_d$ situated a distance $L_d$ from the production region, the number of particles in the decay channel $\{X\}$ that are produced from $\nu_h$ decay \emph{within} the detector region is given by
 \bea & & \Big[  N^{\alpha\,X}_F\Big]_{det} =  N^{\alpha\,X}_F(L_d+\Delta L_d) - N^{\alpha\,X}_F(L_d) \nonumber \\ & = &
  \Bigg[ \Gamma_{\Phi}(k)\,  e^{-\Gamma_h(q^*) \frac{L_d}{v^*_g}} \,
   \frac{\Big(1-e^{-\Gamma_h(q^*) \frac{\Delta L_d}{v^*_g}}\Big)}{\Gamma_{\Phi}(k)- {\Gamma_h(q^*)} }
   -  {\Gamma_h(q^*)} \,  e^{-\Gamma_\Phi(k) \frac{L_d}{v^*_g}} \frac{\Big(1-e^{-\Gamma_\Phi(k) \frac{\Delta L_d}{v^*_g}} \Big)}{\Gamma_\Phi(k)-  {\Gamma_h(q^*)}  }
      \Bigg] \nonumber \\ & \times &  BR(\Phi \rightarrow L^\alpha\,\nu_h)\,BR(\nu_h \rightarrow \{X\})\,,\label{nofld}\eea where $\Gamma_h;\Gamma_\Phi$ are the \emph{total} decay rates.

 This is one of the main results of this study. We emphasize that the second term inside the brackets is a consequence of unitarity as discussed above. This result is general, it is  valid for arbitrary production\footnote{Although we focused on production via pseudoscalar meson decay, the formulation can be straightforwardly adapted to heavy charged lepton decay.} and or decay channels and no assumptions have been made on the total decay rates or branching ratios other than their perturbative nature\footnote{The perturbative nature of the decay rates has been used to argue on the separation of time scales and the validity of the derivative expansion in the Wigner-Weisskopf approximation.}.

 Various limits are of experimental relevance: \emph{if} the position of the detector $L_d\gg v_g \tau_\Phi$ \emph{and} $\Gamma_\Phi > \Gamma_h(q^*)$ the second term in (\ref{nofld}) is subleading and can be neglected, in this case the number of decay products (for one initial meson), simplifies to
 \be \Big[  N^{\alpha\,X}_F\Big]_{det} =  e^{-\Gamma_h(q^*) \frac{L_d}{v^*_g}} \,
   \frac{\Big(1-e^{-\Gamma_h(q^*) \frac{\Delta L_d}{v^*_g}}\Big)}{1- \frac{\Gamma_h(q^*)}{\Gamma_\Phi} }~ BR(\Phi \rightarrow L^\alpha\,\nu_h)\,BR(\nu_h \rightarrow \{X\})\,. \label{nomes}\ee For $\Gamma_h/\Gamma_\Phi \ll 1$ the result (\ref{nomes}) coincides with the number of decays in the detector volume used in ref.\cite{gninenko}\footnote{Identifying $L_d \rightarrow L'$ and $\Delta L_d \rightarrow L$ in ref.\cite{gninenko}.}.   Furthermore,  for $\Gamma_\Phi \gg \Gamma_h$ and \emph{if} $\Gamma_h(q^*) \frac{\Delta L_d}{v^*_g} \ll 1$ \emph{and} $ \Gamma_\Phi \frac{L_d}{v^*_g} \gg 1$ the approximation
\be \Big[  N^{\alpha\,X}_F\Big]_{det} \simeq \Bigg[
 \Gamma_h(q^*) \frac{\Delta L_d}{v^*_g} \Bigg] ~ BR(\Phi \rightarrow L^\alpha\,\nu_h)\,BR(\nu_h \rightarrow \{X\}) \label{approxNX}\ee holds. This approximation has been invoked in refs.\cite{tao,dib} and in the experimental analysis in ref.\cite{exp10}. However, these approximations require that besides $\Gamma_\Phi \gg \Gamma_h$, \emph{both the position of the detector and its fiducial length} $\Delta L_d$ be much smaller than $v^*_g \tau_h(q^*)$. The reliability of these approximations must be assessed on a case by case basis.

 \vspace{2mm}

 \textbf{Generalization:} In the description above we have envisaged that heavy sterile neutrinos are produced in the decay of pseudoscalar mesons, but this assumption can be relaxed and generalized straightforwardly. In particular in ref. \cite{gninenko,gninenko2} it is proposed that heavy sterile neutrinos are produced via neutral current interactions between active neutrinos and nucleons in the detector medium, namely $\nu_\mu + N \rightarrow \nu_h + N'$ with a subsequent radiative decay $\nu_h \rightarrow \nu_a +\gamma$. This situation can be included in the description above as follows: the initial state is now  the two particle initial state of definite total momentum and energy
  $ |\nu_{\mu,\vec{k}-\vec{k}_N}; N_{\vec{k}_N}\rangle  $ rather than a single particle meson state and the intermediate state(s) are of the form $|\nu_{i,\vec{q}};N'_{\vec{k}-\vec{q}}\rangle$ ($i=a,h$)). Therefore we can apply the results obtained above by the replacement
  \be  |\Phi_{\vk}\rangle \rightarrow   |\nu_{\mu,\vec{k}-\vec{k}_N}; N_{\vec{k}_N}\rangle ~~;~~
\big|\nu_{i,\vq};\,L^\alpha_{\vk-\vq} \rangle \rightarrow
|\nu_{i,\vec{q}};N'_{\vec{k}-\vec{q}}\rangle \,, \label{gener}\ee and the decay rate $\Gamma_{\Phi}$ must be replaced by the total transition probability per unit time, namely
\be \Gamma_{\Phi} \rightarrow \sum_{i=a,h}\Gamma_{\nu_\mu \,N \rightarrow \nu_i\, N'} \,. \label{toti}\ee

The equation equation used in ref.\cite{gninenko2} to obtain the expected number of signal events $\nu_h \rightarrow \nu_a \gamma$ within the fiducial volume of the detector implicitly (or explicitly) \emph{assumes} that the total transition rate (\ref{toti}) is \emph{much larger} than $\Gamma_h$.

Whereas this generalization accounts for the production of heavy sterile neutrinos envisaged in ref.\cite{gninenko2}, the experimental search for radiative decay of heavy sterile neutrinos reported in ref.\cite{istra}  involves the cascade decay $K^- \rightarrow \mu^- \nu_h \rightarrow \mu^- \nu_a \gamma$ which is described by the original framework and is discussed below.

\section{Two examples of ``visible'' decay:} We consider two relevant examples of ``visible'' decay: i)$\nu_h \rightarrow e^+ e^- \nu_a$, via a Standard Model charged current vertex, and ii) $\nu_h \rightarrow \nu_a\gamma$ via a transition magnetic moment. As main production mechanism we consider heavy neutrinos produced from $K$ decay at rest $K \rightarrow L^\alpha \nu_h$ for both cases. The experiment reported in ref\cite{istra} precisely searches for radiative decays of heavy sterile neutrinos in the cascade process $K^- \rightarrow \mu^- \nu_h \rightarrow \mu^- \nu_a \gamma$, therefore our study directly addresses this experiment.

 The largest mass window for a heavy sterile neutrino in pseudoscalar decay is available in $K^+ \rightarrow e^+ \nu_h$. For pseudoscalar meson decay at rest,
\be q^* = \frac{M_\Phi}{2}\, \Big[\lambda(1,\delta_{\alpha},\delta_{h}) \Big]^\frac{1}{2}~;~E_h(q^*) = \frac{M_\Phi}{2}\,(1+\delta_h-\delta_\alpha) \label{qEstar}\ee where
\be \lambda(x,y,z) = x^2+y^2+z^2-2xy-2xz-2yz~~;~~ \delta_\alpha  = \frac{m^2_{L^\alpha}}{M^2_\Phi} ~;~\delta_{h} = \frac{m^2_{h}}{M^2_\Phi}\,. \label{dam}\ee

 The ratio between the decay rate of a parent meson $\Phi$ to a heavy $\nu_h$ and   to a \textbf{massless} (flavor) neutrino $\nu_\alpha$ is\cite{shrock}
\be R(\Phi;\alpha;h) = \frac{\Big[\lambda(1,\delta_{\alpha},\delta_{h}) \Big]^\frac{1}{2}\,(\delta_{\alpha}+\delta_{h}-(\delta_{\alpha}-\delta_{h})^2)}{\delta_{\alpha}(1-\delta_{\alpha})^2} \label{ratio} \ee
  therefore
\be \Gamma(\Phi\rightarrow L^\alpha\,\nu_h) = \Gamma_\Phi \,BR(\Phi\rightarrow L^\alpha \nu_\alpha)\,R(\Phi;\alpha;h)\,|U_{\alpha h}|^2 \,.\label{gamanuh}\ee

In $K^+ \rightarrow e^+ \nu_h$ decay at rest,  the group velocity for a wave packet of $\nu_h$ is given by
\be v^*_g = \frac{q^*}{E_h(q^*)} = \frac{\Big[\lambda(1,\delta_{\alpha},\delta_{h})\Big]^{1/2}}{(1+\delta_h-\delta_\alpha)} \,. \label{vgh}\ee

For the $K^+ \rightarrow e^+ \nu_e$ channel  $R(K;e ;h)$ features a broad maximum in the region $0.1 \lesssim \delta_h \lesssim 0.9$ where $R(K;e;h)\simeq 10^5$\cite{shrock} and within this wide mass region $BR(K\rightarrow e \nu_e)\,R(K;e;h) \simeq 1$ and $\Gamma(K \rightarrow e\,\nu_h) \simeq \Gamma_K \, |U_{eh}|^2$, so that $BR(K\rightarrow e\,\nu_h) \simeq |U_{eh}|^2$. The upper bounds obtained in refs.\cite{exp10,kuseexpt2}  and the most recent experimental upper bounds from ref.\cite{exp12} (PS191 experiment at CERN) from $K^+\rightarrow e^+\,\nu_h\rightarrow e^+(e^+ e^-\nu_e)$ ($K_e(1)$ data in fig. (1) in the second reference in\cite{exp12}) yields $10^{-9}\lesssim |U_{eh}|^2 \lesssim 10^{-5}$ in the mass range $50\,\mathrm{MeV} \lesssim m_h \lesssim 400 \,\mathrm{MeV}$, whereas data from the IHEP-JINR neutrino detector\cite{exp10} for the same decay channel reports $10^{-7}\lesssim |U_{eh}|^2 \lesssim 10^{-4}$ within the same mass region (see fig.4 in ref.\cite{exp10}).

 \vspace{2mm}

 \textbf{i:)} $\mathbf{ \nu_h \rightarrow e^+ e^- \nu_a :}$

 \vspace{2mm}

 The decay rate for $\nu_h \rightarrow e^+ e^- \nu_a$ in the rest frame of $\nu_h$ has been obtained in ref.\cite{shrock}, neglecting the mass of  $\nu_a$ it is given by
\be \Gamma(\nu_h \rightarrow e^+ e^- \nu_a) = \frac{G^2_F\, m^5_{h}}{192 \pi^3} ~ \mathcal{H}\Big(\frac{m^2_e}{ m^2_{h}}\Big) \, |U_{eh}|^2  \label{gamahee}\ee
where\cite{shrock}
\be \mathcal{H}(x) = (1-4x^2)^\frac{1}{2}\,(1-14x-2x^2-12x^3)+24x^2(1-x^2)\ln\frac{1+(1-4x^2)^\frac{1}{2}}{1-(1-4x^2)^\frac{1}{2}} \,.\ee Neglecting $m_e$ it follows that
\be \Gamma(\nu_h \rightarrow e^+ e^- \nu_a) \simeq 3.5\times  \, \Bigg[\frac{m_h}{100\,\mathrm{MeV}}\Bigg]^5\,  \Bigg[ \frac{|U_{eh}|^2 }{ 10^{-5}}\Bigg] \,\mathrm{s}^{-1}\,, \label{Ghee}\ee with an extra factor of $2$ if $\nu_h$ is a Majorana neutrino. Assuming that $\nu_h \rightarrow e^+ e^-\nu_a$ has a branching ratio of $\mathcal{O}(1)$  and with $50\,\mathrm{MeV} \lesssim m_h \lesssim 400 \,\mathrm{MeV}$, the constraint  $10^{-7}\lesssim |U_{eh}|^2 \lesssim 10^{-4}$ implies that $\Gamma_h \ll \Gamma_K \simeq 10^{8}\, \mathrm{s}^{-1}$. Within this mass range  $0.2 \lesssim v^*_g \lesssim 0.98$ therefore for $K$ decay at rest and for a detectore placed at $L_d \gg 1\,\mathrm{m}$  the approximations leading to (\ref{approxNX}) are justified, namely

\be \Big[  N^{e^+e^- \nu_{e}}_F\Big]_{det} \simeq \Bigg[
 \Gamma_h(q^*) \frac{\Delta L_d}{v^*_g} \Bigg] ~ BR(K^+ \rightarrow e^+\,\nu_h)\,BR(\nu_h \rightarrow e^+ e^-\nu_e)\,, \ee which for this particular case yields
 \be \Big[  N^{e^+e^- \nu_{e}}_F\Big]_{det} \simeq  3.6 \times10^{-13} \sqrt{\delta_h}\,\Big(\frac{m_h}{100\,\mathrm{MeV}}\Big)^5\,  \Bigg[\frac{|U_{eh}|^2}{10^{-5}}\Bigg]^2\,\Bigg[ \frac{ (\delta_{e}+\delta_{h}-(\delta_{e}-\delta_{h})^2)}{10^5\,\delta_{e}(1-\delta_{e})^2}  \Bigg]\,\Big( \frac{\Delta L_d}{\mathrm{mts}}\Big)\,,\label{finfor} \ee
  where we used $BR(K^+\rightarrow e^+ \nu_e) = 1.55 \times 10^{-5}$. A similar analysis can be carried out for any other decay channel. The total number of detected lepton pairs is obtained by multiplying by the total number of mesons in the beam and the detector acceptance and reconstruction efficiency etc.

 \vspace{2mm}

 \textbf{ii:)} $\mathbf{ \nu_h \rightarrow \gamma\, \nu_a :}$

 \vspace{2mm}

 In ref.\cite{gninenko} it was argued that the LSND and MiniBooNE  anomalies, an excess of electron-like events in quasi-elastic charged current events  may be explained by the radiative decay of a heavy sterile neutrino  via an anomalous transition magnetic moment where the heavy sterile neutrino $\nu_h$ is produced at LSND/MiniBooNE in a $\mu$ charged current vertex. The analysis of ref.\cite{gninenko} suggests that the anomalies may be explained by this decay channel of the heavy neutrino if
 \be 40\,\mathrm{MeV} \lesssim m_h \lesssim 80 \, \mathrm{MeV}~;~10^{-3} \lesssim |U_{\mu h}|^2 \lesssim 10^{-2}~;~ 10^{-11}\,\mathrm{s} \lesssim \tau_h \lesssim 10^{-9}\,\mathrm{s} \,. \label{raddec}\ee although this parameter space has been found in tension by the analysis in ref.\cite{dibfot} and the experimental results  on searches of radiative decays from $K^-\rightarrow \mu^-\nu_h\rightarrow (\mu^-)\nu_a\gamma$ at ISTRA in ref.\cite{istra}. Although it is argued in ref.\cite{gninenko2} that the tension may be alleviated by a suppression of the $\nu_h$ charged current channels.

 For the production channel $K^+\rightarrow \mu^+ \nu_h$ the analysis of ref.\cite{shrock} shows that $R(K,\mu;h)$ features a broad maximum in the region $0.1 \lesssim (m_h/M_K-m_\mu)^2 \lesssim 0.8$ with $ 1 \lesssim R(K,\mu;h) \lesssim 5$ again leading to $BR(K \rightarrow \mu \nu_\mu)\,R(K,\mu;h) \simeq \mathcal{O}(1)$ within this mass range for $\nu_h$.

 In ref.\cite{gninenko,gninenko2} it is argued that the radiative decay $\nu_h \rightarrow \nu_a\,\gamma$ may be  mediated by a transition anomalous magnetic moment and that in order for the radiative decay of a heavy sterile neutrino to explain the low energy enhancement of electron-like events at MiniBooNE the branching ratio for this process must be nearly unity, although a recent different analysis suggests a much smaller branching ratio\cite{masip}.

  The range for the $\nu_h$ lifetime (\ref{raddec}) argued in these references $10^{-11}\,\mathrm{s} \lesssim \tau_h \lesssim 10^{-9}\,\mathrm{s} $ is \emph{much shorter} than the lifetime of the parent meson $\tau_K = 1.24\times 10^{-8}\,\mathrm{s}$,  and this situation entails important corrections.

 When $\Gamma_h \geq \Gamma_K$ as is suggested by the analysis in refs.\cite{gninenko,gninenko2} the full expression (\ref{NXs}) must be considered, which when combined with the wave packet analysis above yields for the number of photons in the final state as a function of the distance $L$ from the production region (assuming $K$ decay at rest)
 \be N^\gamma(L) = \Bigg[\frac{\Gamma_K\,\Big(1-e^{-\Gamma_h(q^*) \frac{L}{v^*_g}}\Big) - \Gamma_h(q^*)\,\Big(1-e^{-\Gamma_K  \frac{L}{v^*_g}} \Big)}{ \Gamma_K- \Gamma_h(q^*) } \Bigg]\,BR(K^+ \rightarrow \mu^+\,\nu_h)\,BR(\nu_h \rightarrow \nu_a \gamma)\,.\label{numfoti} \ee

  In comparison, the expression used for the analysis of the MiniBooNE/LSND data in ref.\cite{gninenko} corresponds to
 \be N^\gamma_G(L) =  \Big(1-e^{-\Gamma_h(q^*) \frac{L}{v^*_g}}\Big) \,\,BR(K^+ \rightarrow \mu^+\,\nu_h)\,BR(\nu_h \rightarrow \nu_a \gamma)\,. \label{nfotgni}\ee

  \begin{figure}[h!]
\begin{center}
\includegraphics[height=3.2in,width=3.2in,keepaspectratio=true]{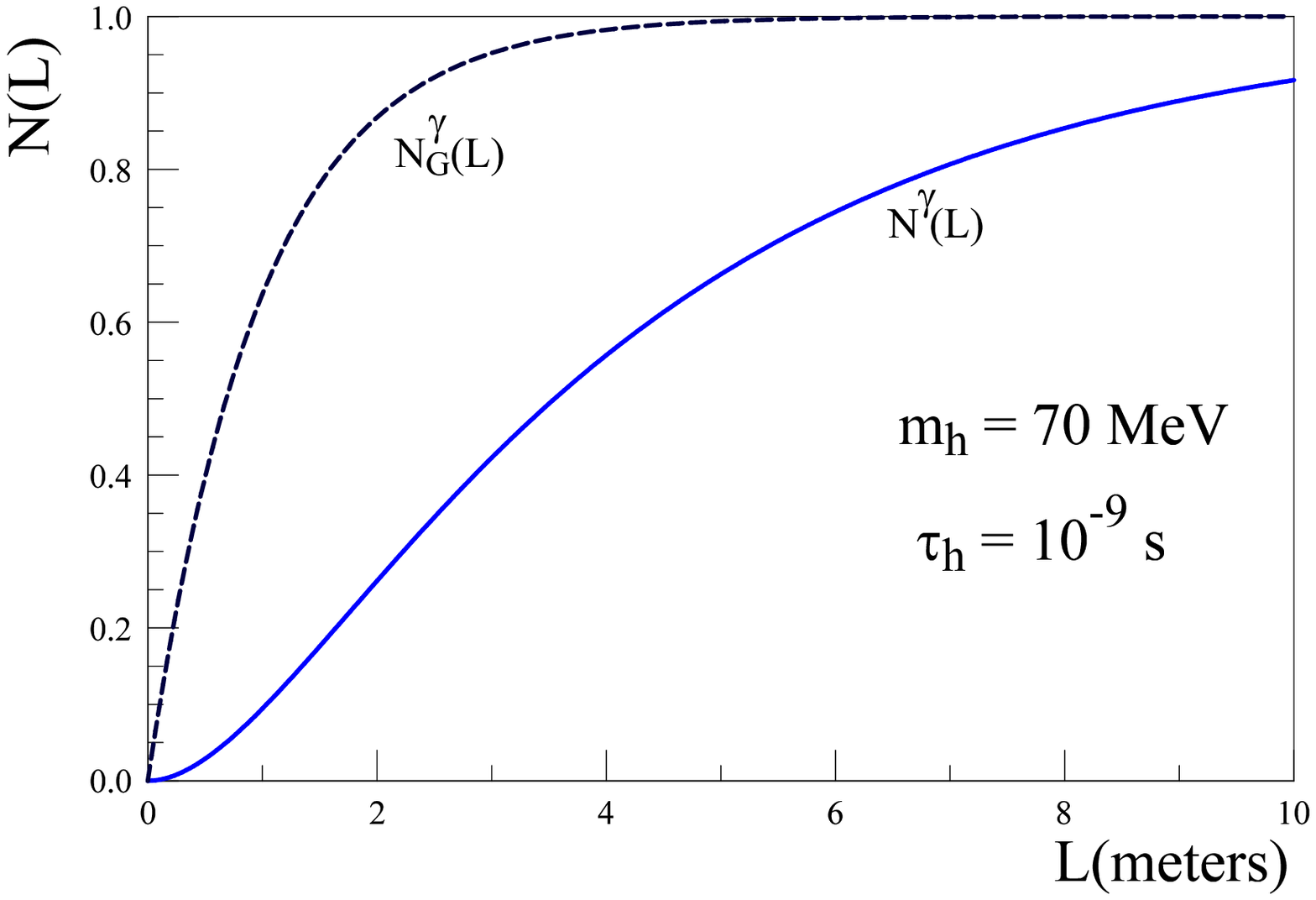}
\includegraphics[height=3.2in,width=3.2in,keepaspectratio=true]{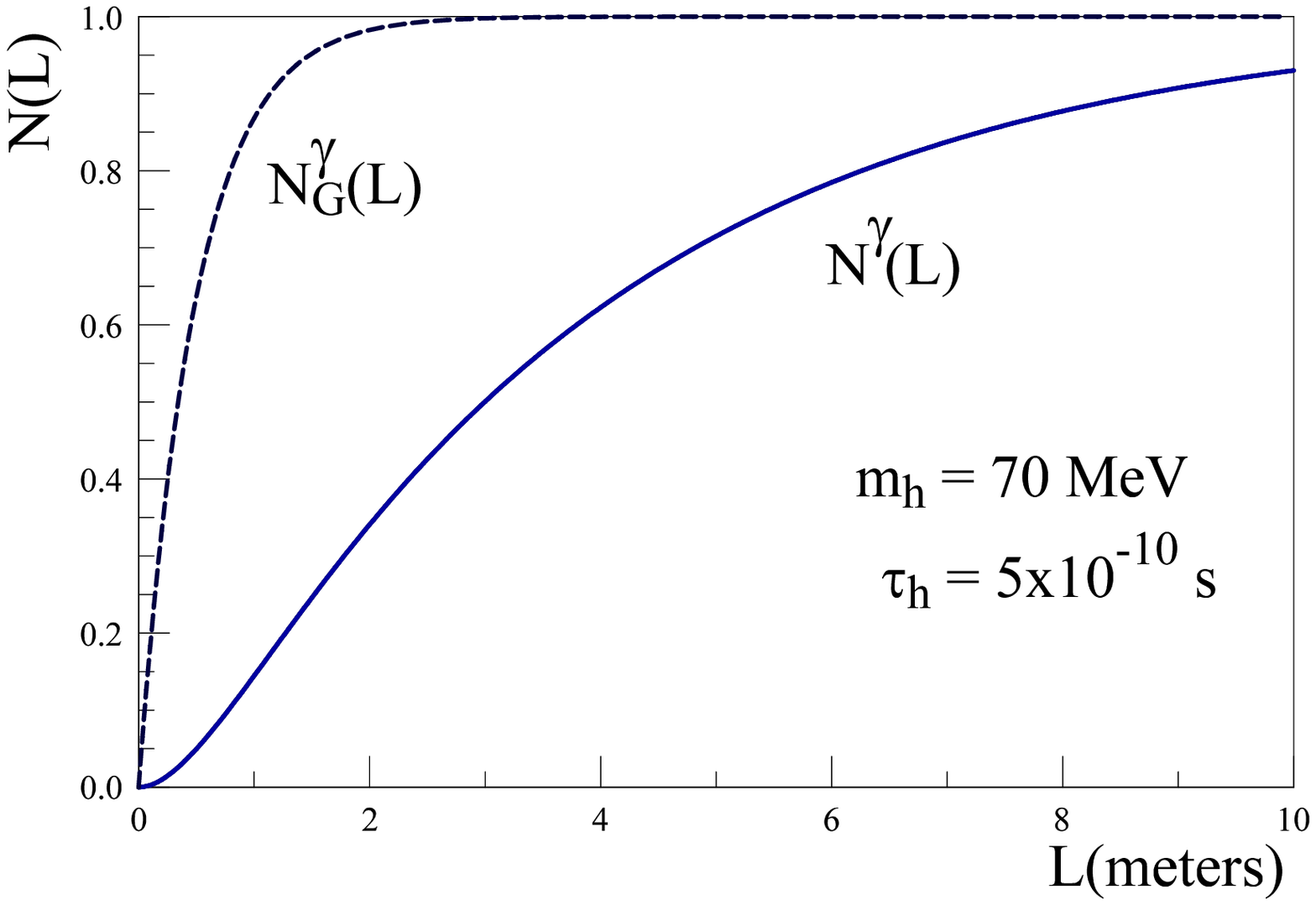}
\caption{$N^\gamma(L)$ eqn. (\ref{numfoti}) and $N^\gamma_G(L)$ eqn. (\ref{nfotgni}) for $m_h=70\,\mathrm{MeV}~;~\tau_h=10^{-9},5\times 10^{-10}$ respectively for $K$ decay at rest and $BR(K^+ \rightarrow \mu^+\,\nu_h)\,BR(\nu_h \rightarrow \nu_a \gamma)=1$. }
\label{fig:numfot}
\end{center}
\end{figure}

Fig. (\ref{fig:numfot}) shows (\ref{numfoti}) and (\ref{nfotgni})  for $m_h=70\,\mathrm{MeV}~;~\tau_h=10^{-9},5\times 10^{-10}\,\mathrm{s}$ respectively for $K$ decay at rest and $BR(K^+ \rightarrow \mu^+\,\nu_h)\,BR(\nu_h \rightarrow \nu_a \gamma)=1$ displaying a substantial difference in the number of photons  produced at a distance $L$ from the production region.

The number of photons produced within the fiducial length $\Delta L_d$ of a detector placed at a distance $L_d$ from the production region is given by (\ref{nofld}), namely
 \bea \Big[  N^{\gamma}\Big]_{det} & =  &
  \Bigg[   \Gamma_K\, e^{-\Gamma_h(q^*) \frac{L_d}{v^*_g}} \,
   \frac{\Big(1-e^{-\Gamma_h(q^*) \frac{\Delta L_d}{v^*_g}}\Big)}{\Gamma_K- \Gamma_h(q^*)  }
   -  {\Gamma_h(q^*)} \,  e^{-\Gamma_K \frac{L_d}{v^*_g}} \frac{\Big(1-e^{-\Gamma_K \frac{\Delta L_d}{v^*_g}} \Big)}{\Gamma_K-  {\Gamma_h(q^*)}  }
      \Bigg] \nonumber \\ & \times &  BR(K^+ \rightarrow \mu^+\,\nu_h)\,BR(\nu_h \rightarrow \nu_a \gamma)\,,\label{numfotLd}\eea and the corresponding quantity obtained from (\ref{nfotgni}) is
  \be  \Big[  N^{\gamma}_G\Big]_{det} =     e^{-\Gamma_h(q^*) \frac{L_d}{v^*_g}} \,
    {\Big(1-e^{-\Gamma_h(q^*) \frac{\Delta L_d}{v^*_g}}\Big)}BR(K^+ \rightarrow \mu^+\,\nu_h)\,BR(\nu_h \rightarrow \nu_a \gamma)\,, \label{ndetgni} \ee which is precisely the quantity used in refs.\cite{gninenko,gninenko2}.

  \begin{figure}[ht!]
\begin{center}
\includegraphics[height=3.2in,width=3.2in,keepaspectratio=true]{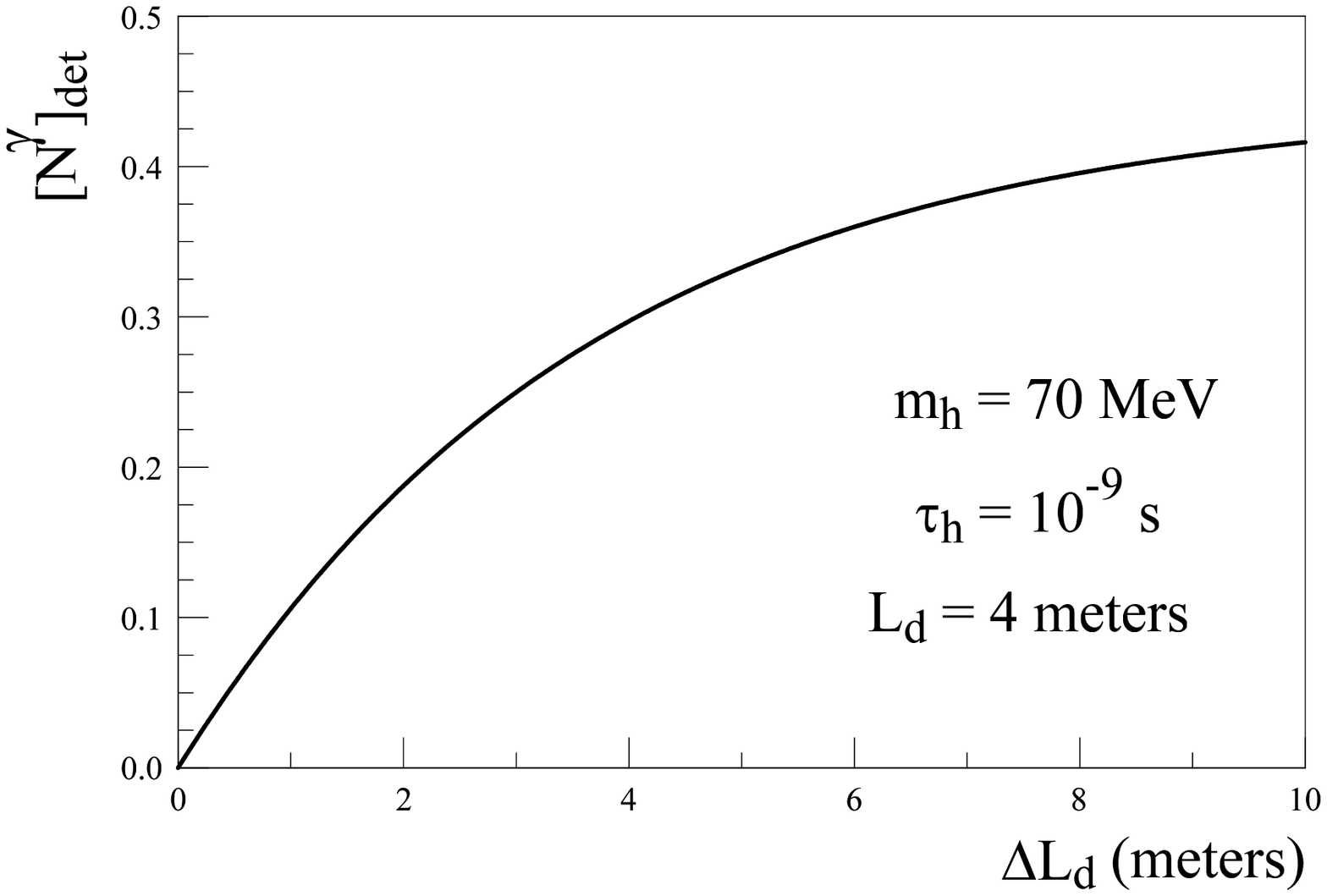}
\includegraphics[height=3.2in,width=3.2in,keepaspectratio=true]{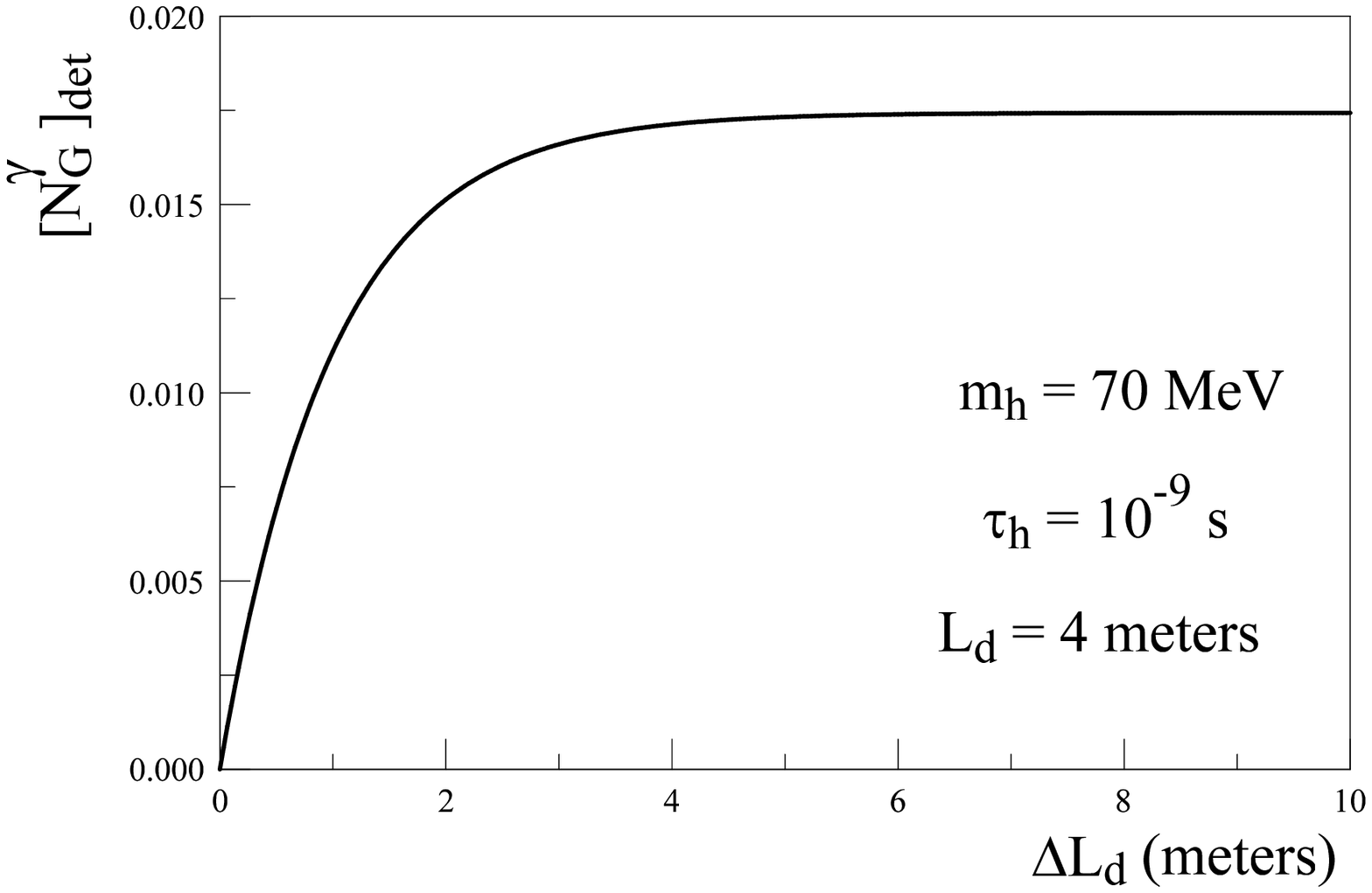}
\includegraphics[height=3.2in,width=3.2in,keepaspectratio=true]{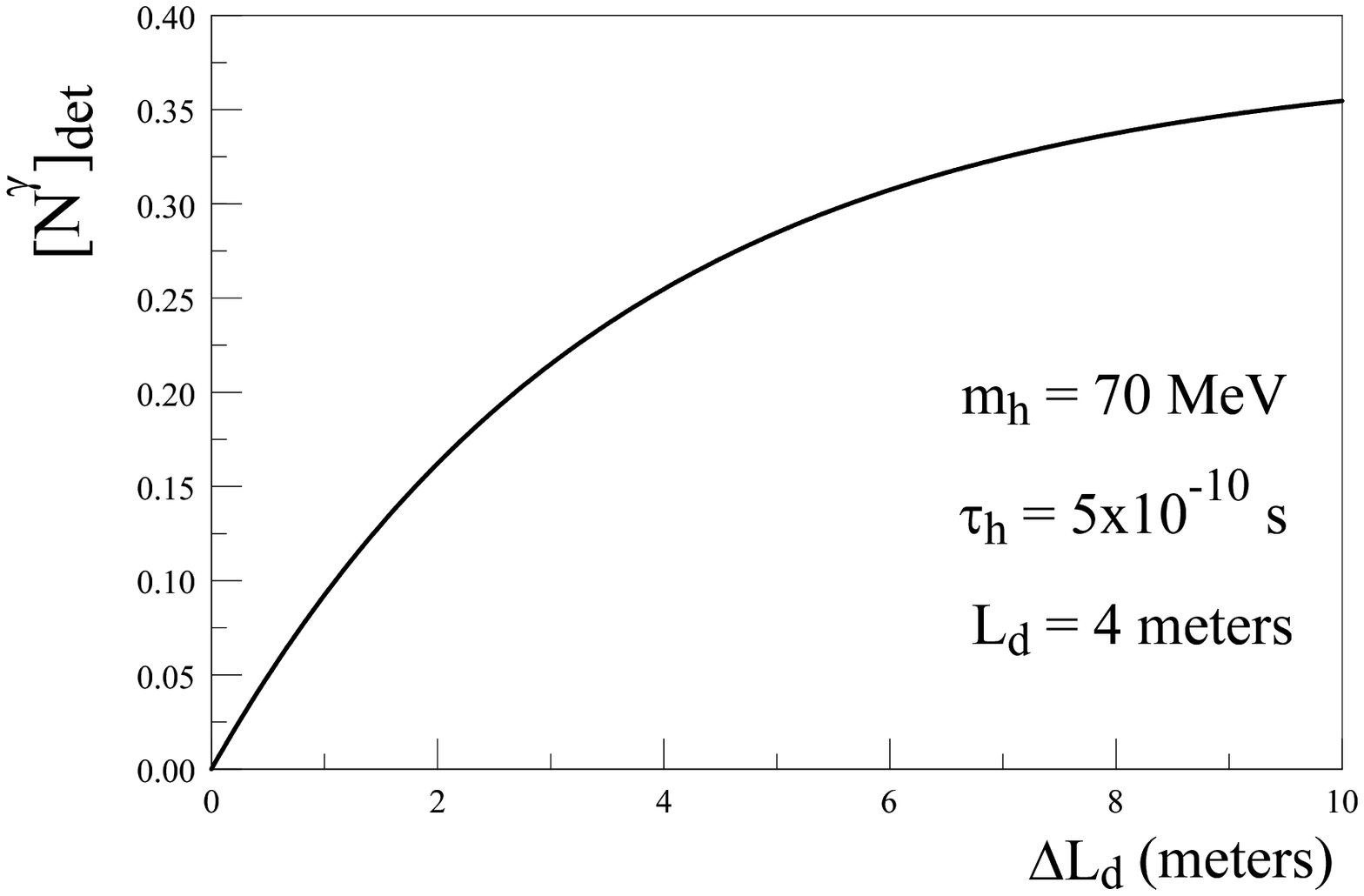}
\includegraphics[height=3.2in,width=3.2in,keepaspectratio=true]{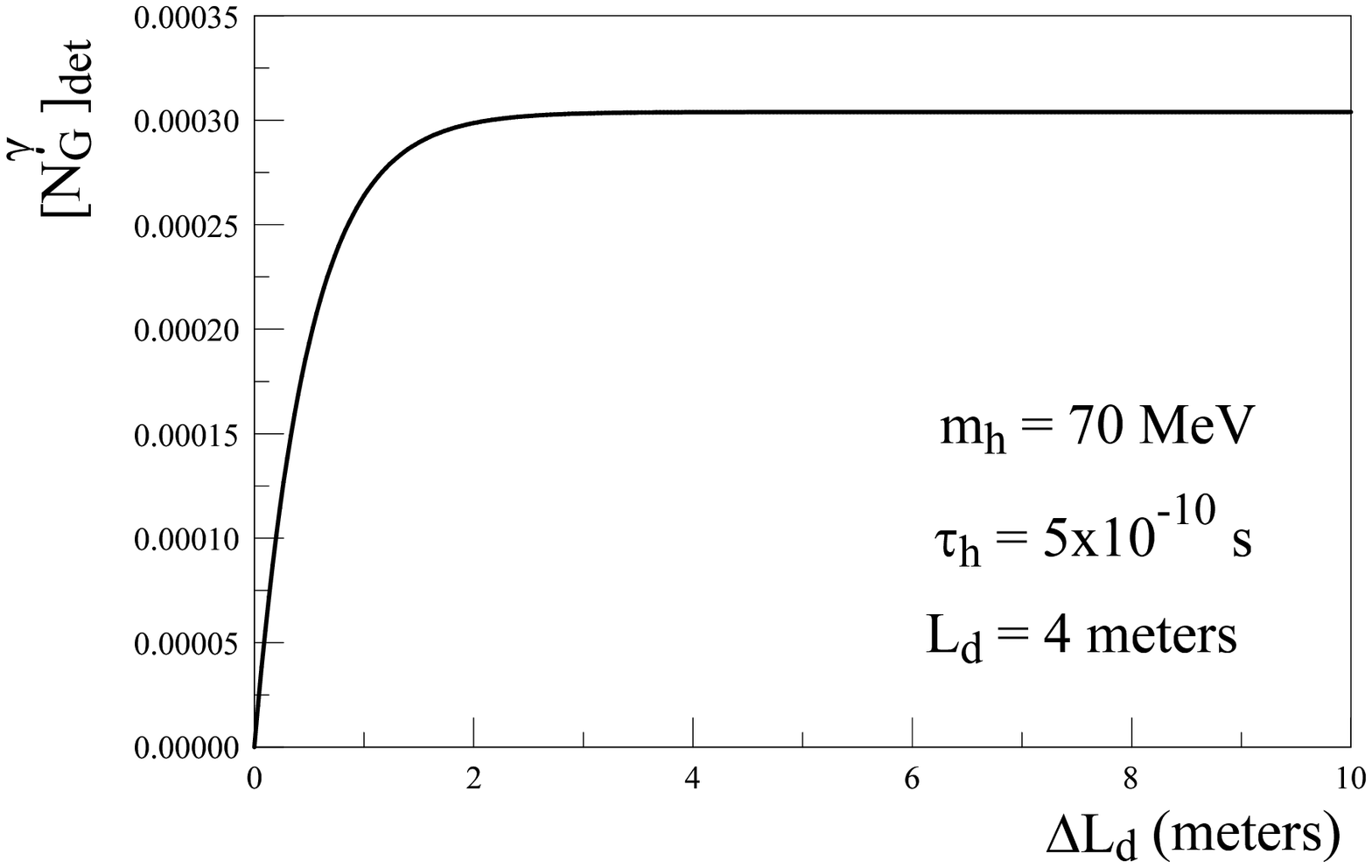}
\caption{$\Big[  N^{\gamma}\Big]_{det} $ eqn. (\ref{numfotLd}) and $\Big[  N^{\gamma}_G\Big]_{det}$ eqn. (\ref{ndetgni}) for $m_h=70\,\mathrm{MeV}~;~\tau_h=10^{-9},5\times 10^{-10}\,\mathrm{s}$ respectively for $K$ decay at rest and $BR(K^+ \rightarrow \mu^+\,\nu_h)\,BR(\nu_h \rightarrow \nu_a \gamma)=1~;~L_d=4\,\mathrm{mts}$. }
\label{fig:numfotdet}
\end{center}
\end{figure}

 Fig.(\ref{fig:numfotdet}) shows (\ref{numfotLd}) and (\ref{ndetgni}) for $m_h=70\,\mathrm{MeV}~;~\tau_h=10^{-9},5\times 10^{-10}\,\mathrm{s}$ respectively for $K$ decay at rest and $BR(K^+ \rightarrow \mu^+\,\nu_h)\,BR(\nu_h \rightarrow \nu_a \gamma)=1$ for $L_d = 4\,\mathrm{mts}$ which is approximately the decay length of Kaons at rest. These figures show dramatic differences between the expressions, by several orders of magnitude. The main difference can be traced to the contribution from the second term inside the bracket in (\ref{numfotLd}), as emphasized in the previous section, this term is manifestly a consequence of  unitarity.

 This analysis suggests that  in the case advocated in ref.\cite{gninenko,gninenko2} as a possible explanation of the MiniBooNE/LSND anomalies, the space-time evolution of sterile neutrinos brings important modifications in the assessment of the production of final states in the detector region. These important differences may relieve the tension between the parameter region discussed in \cite{gninenko,gninenko2} the theoretical analysis of ref.\cite{dibfot} and the results from the ISTRA experiment\cite{istra}, although the analysis presented in ref.\cite{istra} does not seem to include the spatio-temporal contribution that is exponentially suppressed in the distances.

\section{Summary and conclusions.}

Several well motivated extensions of the Standard Model predict the existence of heavy  ``sterile'' neutrinos that mix with the active neutrinos with small mixing matrix elements. These heavy sterile neutrinos may play an important role in   astrophysical and cosmological phenomena, in baryogenesis through leptogenesis,   could be suitable dark matter candidates and may also   explain anomalous low energy events at MiniBooNE and LSND via their radiative decay. If these heavy sterile neutrinos are Majorana they mediate lepton number violating transitions with potentially observable effects. These aspects motivate a rekindling of efforts to search for heavy sterile neutrinos in current and future experimental facilities through various recent proposals. Most current bounds on the mixing matrix elements between a heavy sterile $\nu_h$ and active neutrinos suggest that $|U_{eh}|^2~;~|U_{\mu h}|^2 \lesssim 10^{-7}-10^{-5}$ in the mass range $30\,\mathrm{MeV} \lesssim m_h \lesssim 400\,\mathrm{MeV}$ rendering production and decay rates of heavy sterile neutrinos exceedingly small and resulting in displaced vertices. If kinematically allowed, heavy sterile neutrinos may be produced as resonant states for example in the decay of pseudoscalar mesons, resulting in transition matrix elements that are resonantly enhanced. As they are produced ``on-shell'' they propagate over large distances prior to decaying into a set of final states $\{X\}$ either away from the detector, or within the fiducial volume of detectors placed far away from the production region. This is a cascade decay process $\Phi \rightarrow L \nu_h \rightarrow (L) \{X\}$.

There are two main ingredients required to assess their detection via the final decay products: i) an analysis of the production and decay rates and ii) an analysis of the space-time evolution from production to decay.

 While there has been an intense theoretical program to assess production and decay in various channels, much less has been studied with regard to the spatio-temporal evolution that includes within the same framework the dynamics of production, propagation and decay.

  In this article we focused on these latter aspects by implementing a non-perturbative and manifestly unitary quantum field theoretical framework that yields the time dependent amplitudes of intermediate and final states in a cascade decay process. Combined with a wave packet description we obtain a complete spatio-temporal description of the dynamics of the cascade from  production to decay.

We focus the discussion on the generic cascade decay process $\Phi \rightarrow L^\alpha \nu_h~;~ \nu_h \rightarrow \{X\}$ where $\Phi$ is a pseduscalar meson, $L^\alpha$ a charged lepton and $\{X\}$ denotes a generic decay channel for $\nu_h$. Our approach is completely general, it is manifestly unitary and    only inputs   total decay rates for production and decay along with   branching ratios for particular channels. We have provided a generalization of the framework to the case in  which sterile neutrinos are produced via neutral current interactions with nucleons, and further generalizations are straightforward.

Our main result for the number of $\{X\}$ final state particles detected at time $t$ is given by

 \bea  N^{\alpha X}_F (t) & = &    \Bigg[\frac{\Gamma_\Phi(k)\,\Big(1-e^{-\Gamma_h(q^*) t}\Big) - \Gamma_h(q^*)\,\Big(1-e^{-\Gamma_\Phi(k) t} \Big)}{ \Gamma_\Phi(k)- \Gamma_h(q^*) } \Bigg] \nonumber \\ &\times &  BR(\Phi \rightarrow L^\alpha\,\nu_h)\,BR(\nu_h \rightarrow \{X\})\,  \eea where $\Gamma_\Phi(k)~;~\Gamma_h(q^*)$ are the \emph{total} decay rates for the pseudoscalar meson and heavy sterile neutrino respectively and $q^*$ is the value of the momentum with which the heavy neutrinos are produced. This expression is a manifestation of unitary time evolution. Combining this time evolution with a wave packet description of the parent particle we obtain the following expression for the number of particles in a given decay channel that are detected within a fiducial length $\Delta L_d$ of a detector placed a distance $L_d$ away from the production region

 \bea   \Big[  N^{\alpha\,X}_F\Big]_{det}    & = &
  \Bigg[    \Gamma_\Phi(k) \, e^{-\Gamma_h(q^*) \frac{L_d}{v^*_g}} \,
   \frac{\Big(1-e^{-\Gamma_h(q^*) \frac{\Delta L_d}{v^*_g}}\Big)}{\Gamma_\Phi(k)- {\Gamma_h(q^*)} }   -  {\Gamma_h(q^*)} \,  e^{-\Gamma_\Phi(k) \frac{L_d}{v^*_g}} \frac{\Big(1-e^{-\Gamma_\Phi(k) \frac{\Delta L_d}{v^*_g}} \Big)}{\Gamma_\Phi(k) -  {\Gamma_h(q^*)}  }
      \Bigg] \nonumber \\ & \times &  BR(\Phi \rightarrow L^\alpha\,\nu_h)\,BR(\nu_h \rightarrow \{X\})\,, \label{fullexp}  \eea where $v^*_g$ is the group velocity of the heavy sterile neutrino.

      We studied in detail two examples of ``visible'' decay: i) $K^+\rightarrow e^+\nu_h; \nu_h \rightarrow e^+ e^- \nu_a$ via a standard model charged current vertex, and ii)$K^+\rightarrow \mu^+ \nu_h; \nu_h \rightarrow \nu_a \gamma$ the radiative decay of the heavy sterile neutrino being mediated by a  transition magnetic moment, in both cases we consider $K$ decay at rest. The second example of cascade decay has been argued to be a potential explanation of the low energy anomalous electron-like events at MiniBooNE and LSND if the lifetime $10^{-11} \lesssim \tau_h \lesssim 10^{-9}\,\mathrm{s}$ and $BR(\nu_h \rightarrow \nu_a \gamma) \simeq 1$.

      In the first case $\Gamma_K \gg \Gamma_h$ and for $L_d \gtrsim 1\,\mathrm{mt}$ we find for the number of $e^+ e^-$ pairs detected within $\Delta L_d$

      \be \Big[  N^{e^+e^- \nu_{e}}_F\Big]_{det} \simeq \Bigg[
 \Gamma(\nu_h \rightarrow e^+ e^- \nu_e) \frac{\Delta L_d}{v^*_g} \Bigg] ~ BR(K^+ \rightarrow e^+\,\nu_h)\,.  \ee

 In the second case $K^+\rightarrow \mu^+ \nu_h; \nu_h \rightarrow \nu_a \gamma$  the analysis of ref.\cite{gninenko,gninenko2} suggests a heavy sterile neutrino lifetime much \emph{smaller} than that of the parent meson and   the full expression (\ref{fullexp}) is necessary in the analysis. Within the parameter range argued in refs.\cite{gninenko,gninenko2} we find substantial corrections from the space-time evolution which \emph{may} help alleviate the tension with the experimental results reported in ref.\cite{istra}.

The results obtained in this article are quite general, the framework is manifestly unitary and yields a consistent description of the space-time evolution from production to decay. These results complement  the robust theoretical program  assessing decay rates and branching ratios of heavy sterile neutrinos and in  combination with them it provides a consistent  theoretical framework for the analysis and interpretation of the next generation of proposed experiments.

\acknowledgements{The author thanks A. Kusenko for enlightening conversations and acknowledges partial support from NSF-PHY-1202227.}

\end{document}